\documentclass[]{aa}
\usepackage{graphicx,subfigure,natbib,url,txfonts,amsmath, mathtools}
\usepackage{lscape}
\bibpunct{(}{)}{;}{a}{}{,}

\begin{document}

\title{Photometric brown-dwarf classification} 

\subtitle{II. A homogeneous sample of 1361 L and T dwarfs brighter than
  $J=17.5$ with accurate spectral types}

\author{Skrzypek, N.\inst{\ref{inst1}}, Warren,
  S. J.\inst{\ref{inst1}}, Faherty, J.K.\inst{\ref{inst2}}}
\authorrunning{Skrzypek, N., et al.}

\institute{Astrophysics Group, Imperial College London, Blackett
  Laboratory, Prince Consort Road, London SW7 2AZ, UK \label{inst1}
  \and Department of Terrestrial Magnetism, Carnegie Institution of
  Washington, Washington, DC 20015, USA \label{inst2}}

\date{Received <date> / Accepted <data>}

\abstract{We present a homogeneous sample of 1361 L and T dwarfs
  brighter than $J=17.5$ (of which 998 are new), from an effective
  area of 3070 deg$^2$, classified by the {\em photo-type} method to
  an accuracy of one spectral sub-type using $izYJHKW1W2$ photometry
  from SDSS+UKIDSS+WISE. Other than a small bias in the early L
    types, the sample is shown to be effectively complete to the magnitude limit,
    for all spectral types L0 to T8. The nature of the bias is an
    incompleteness estimated at $3\%$ because peculiar blue L dwarfs
    of type L4 and earlier are classified late M. There is a
    corresponding overcompleteness because peculiar red (likely young)
    late M dwarfs are classified early L. Contamination of the sample
    is confirmed to be small: so far spectroscopy has been
obtained for 19 sources in the catalogue and all are confirmed to be
ultracool dwarfs. We provide coordinates and
  $izYJHKW1W2$ photometry of all sources. We identify an apparent
  discontinuity, $\Delta m\sim 0.4$\,mag., in the $Y-K$ colour between
  spectral types L7 and L8. We present near-infrared spectra of nine
  sources identified by {\em photo-type} as peculiar, including a new
  low-gravity source ULAS J005505.68+013436.0, with spectroscopic
  classification L2$\gamma$. We provide revised $izYJHKW1W2$ template
  colours for late M dwarfs, types M7 to M9.}


\maketitle

\section{Introduction}
The discovery of ultracool dwarfs later than spectral type M9 has
proceeded rapidly over the past two decades, resulting in the
definiton of three new spectral classes; successively cooler, the L
\citep{kirkpatrick99, martin99}; T
\citep{geballe02,burgasser02,burgasser06b}; and Y dwarfs
\citep{cushing11}. The current paper focuses on L and T dwarfs. Brown
dwarfs are defined as objects too low in mass to sustain hydrogen
burning in their cores. They cool with age at a rate dependent on mass
\citep{burrows97}, so for a given spectral type there is an age-mass
degeneracy. Objects with spectral types beyond the end of the stellar
main sequence, i.e. $>$L3, are unambiguously brown dwarfs.  Early-type
L dwarfs, $\leq$L3, are a mix of young brown dwarfs and low-mass main
sequence stars.

The study of L and T dwarfs has moved beyond the exploratory stage to
the detailed characterisation of the population by e.g. the
measurement of the luminosity and mass functions
\citep{cruz07,pinfield08,reyle10,burningham10,kirkpatrick12},
kinematics \citep{faherty09,faherty12,reiners09,schmidt10,seifahrt10},
the frequency of close binaries and wide companions
\citep{burgasser06a,burgasser07,faherty10,faherty11,luhman12,deacon14},
and the study of rare types
\citep{burgasser03sub,folkes07,looper08,gizis12,faherty13,liu13}.

The coverage of the LT sequence of large homogeneous samples\footnote{by which we mean samples
  with high completeness and for which the incompleteness is
  accurately quantified} suitable
for statistical analysis is patchy. This is mostly because of the time
required for spectroscopy, but also because the selection methods,
using colour cuts, pick out only a limited range of spectral
types. The largest existing sample of dwarfs in the LT range is the
catalogue of 484 L dwarfs of \citet{schmidt10}, from the Sloan Digital
Sky Survey (SDSS; \citealt{york00}), selected by $i-z$ colour, and
observed within the SDSS spectroscopic campaign. Of these, 460,
i.e. $95\%$, are classified L3 or earlier. The largest existing sample
of T dwarfs, totaling 176 sources, comes from the WISE team, and is
catalogued in the papers of \citet{kirkpatrick11} and
\citet{mace13}. This sample again represents only a narrow range of
spectral types, with 153 classified T5 or later. Furthermore
spectroscopy is incomplete, so the sample only provides a lower limit
to the space density of late-type T dwarfs.

In a previous paper, \citet{skrzypek15}, hereafter Paper I, we
presented a method, named {\em photo-type}, to identify and accurately
classify samples of L and T dwarfs from multi-band photometry alone,
without the need for spectroscopy. The motivation for developing the
method was the need for a much larger homogeneous sample of L and T
dwarfs, spanning the full range of spectral types, from L0 to T8, in
order to characterise the LT population more precisely, by reducing
the statistical errors on the measurements of properties of
interest. As listed in Paper I, such properties include the
  luminosity function, the disk scale height, the frequency of
  binarity, and the population kinematics. A large sample will also
allow a search for rare types, and the discovery of more benchmark
systems i.e. ultracool dwarf companions to stars with measureable
distance and metallicity e.g. \citet{smith14b, smith14}. The method
might also be useful in identifying distant L and T dwarfs in deep
photometric catalogues, which are so faint that spectroscopy would be
difficult.

The {\em photo-type} method works by comparing the multiwavelength
spectral energy distributions (SEDs) of candidates from broadband
photometry, against a set of template SEDs derived by fitting
polynomials to plots of colour against spectral type, using a set of
spectroscopically classified L and T dwarfs. In Paper I we showed that
the method classifies normal sources to an accuracy of one spectral
sub-type r.m.s., and so is competitive with
spectroscopy. Contamination should be very low because the most likely
contaminants, M dwarfs and reddened quasars, are easily discriminated
against.

We have applied {\em photo-type} to an 8-band photometric catalogue,
combining data from SDSS, the UKIRT Infrared Deep Sky Survey (UKIDSS;
\citealt{UKIDSS}), and the Wide-field Infrared Survey Explorer (WISE;
\citealt{WISE}), over 3344\,deg$^2$. The current paper presents the
resulting sample of 1361 sources, brighter than $J=17.5$, comprising
1281 L dwarfs and 80 T dwarfs. This is the largest existing
homogeneous sample of L and T dwarfs. The sample appears to be highly
complete across the full spectral range L0-T8, judged by the success
in rediscovering virtually all known L and T dwarfs in the area
surveyed. We quantify the completeness in this paper. As expected,
contamination is found to be low: so far spectroscopy has been
obtained for 19 sources in the catalogue and all are confirmed to be
ultracool dwarfs.

The current paper is the companion to Paper I, and follows
  directly from it. Paper I described the motivation for the {\em
    photo-type} method, and the method itself, and quantified the
  accuracy of the {\em photo-type} classifications. The current paper
  presents the sample of L and T dwarfs derived using the {\em
    photo-type} method, and quantifies the completeness of the
  sample. The layout of the remainder of the paper is as follows. In
\S\ref{paper1} we provide a brief summary of Paper I, describing the
      {\em photo-type} method, and its application to the
      SDSS+UKIDSS+WISE catalogue. We also describe two minor updates
      to the method there. \S\ref{sample} presents the new sample of
      1361 sources, and summarises the main characteristics of the
      sample. In \S\ref{completeness} we establish the completeness of
      the sample, considering several possible sources of
      incompleteness.  \S\ref{spectra} presents confirmation spectra
      of 11 objects from the sample, mostly selected by large
      $\chi^2$.  \S\ref{summary} provides a summary of the paper.

\section{Sample selection by {\em photo-type}}
\label{paper1}

The photometric bands used in this study are the $i$ and $z$ bands in
SDSS, the $Y$, $J$, $H$, $K$ bands in UKIDSS, and the $W1$, $W2$ bands
in WISE. All the magnitudes and colours quoted in this paper are
  Vega based. The $YJHKW1W2$ survey data are calibrated to Vega,
while SDSS is calibrated on the AB system; We have applied the offsets
tabulated in \citet{hewett06} to convert the SDSS $iz$ AB magnitudes
to Vega.

Paper I describes the {\em photo-type} method, and the creation of a
multi-band catalogue of point sources combining SDSS, UKIDSS, and WISE
data over 3344\,deg$^2$, used for the search for L and T dwarfs. Here
we summarise details from Paper I relevant to understanding the
contents of the new sample of L and T dwarfs.

The method {\em photo-type} works by comparing the SED of a source,
from multiband photometry, against a library of template SEDs. These
include L and T dwarfs, of all spectral types, as well as quasars,
white dwarfs, and main sequence dwarfs O-M. Attention is restricted to
the colour range $Y-J>0.8$, so that in practice the only contaminants
of the sample of L and T dwarfs are M dwarfs and reddened quasars.

The template SEDs are defined by fitting polynomials to plots of
colour against spectral type, for the 7 colours $i-z$, $z-Y$, $Y-J$,
$J-H$, $H-K$, $K-W1$, $W1-W2$, using the measured colours of 190 L and
T dwarfs with spectroscopic classifications. By anchoring to $J=0$ the
colours define the template SED for each spectral type, L0 to T8. The
best fit template to the SED (i.e. the multiband photometry) of any
target is found by calculating, for each template, the magnitude
offset that minimises the $\chi^2$ of the fit to the SED, and
selecting the template with the minimum value of minimum $\chi^2$. In
making the fit to any SED, the error on each point includes two
contributions: the random photometric error, and an additional error
of 0.05\,mag. per band that accounts for the intrinsic spread in
colours of the population. The errors on the polynomial fits
themselves are also relevant, but were found not to contribute
significantly to the uncertainty in the classification.

The classification of L dwarfs is tied to the optical system of
\citet{kirkpatrick99} and the classification of T dwarfs is tied to
the near-infrared system of \cite{burgasser06b}. It is important to be
clear what this means. The {\em photo-type} method is not designed to
get as close as possible to the standard (optical for L, near-infrared
for T) spectroscopic classification. Rather, it matches the
multiwavelength SED ($0.75-4.6\mu$m) against template SEDs that are
averages for normal L and T dwarfs that have been classified
spectroscopically. For normal objects we can expect the {\em
  photo-type} classification to match the standard spectral
classification closely. For peculiar objects this will not be the
case. For example, L dwarfs that are peculiarly blue for their
spectral type (e.g. subdwarfs) will have an earlier {\em photo-type}
classification than the spectroscopic classification (and {\em vice
  versa} for red objects). This is because {\em photo-type} uses
colours, whereas the spectroscopic classification uses absorption
features, independent of colour. We quantify this bias in
\S\ref{peculiarblue}.

While {\em photo-type} yields biased spectral types for peculiar blue
and red sources, in recompense it has the advantage over spectral
classification of the broad wavelength coverage. Many peculiar
sources, classified as normal with spectroscopy covering a limited
wavelength range, can be recognised as peculiar by the high $\chi^2$
of the {\em photo-type} fit. For example an unresolved LT binary might
be classified differently from an optical or a near-infrared spectrum,
but as normal in both cases. It would therefore require both spectra
to recognise the source as peculiar. The {\em photo-type}
classification would likely be somewhere in between the two spectral
classifications, but the source would be recognised as peculiar by the
high $\chi^2$ of the fit. The $\chi^2$ distribution of the sample is
discussed in \S3. In summary {\em photo-type} provides accurate
classifications for normal sources. For peculiar sources, including
subdwarfs, very red sources, and binaries, the {photo-type} and
standard spectral classifications may not agree, but {\em photo-type}
identifies peculiar sources by their high $\chi^2$. Overall, recognising
that multiwavelength photometry is a measurement of the spectrum at
very low resolution, we can see that spectroscopic classification has
the advantage of much higher resolution, while {\em photo-type} has the
advantage of much broader wavelength coverage.

In deriving the templates the assumption was made that the sample of
known sources is representative of the distribution of colours of the
L and T population, so that the average colours are not
biased. Because the SEDs of the contaminating population, reddened
quasars, are so different from the SEDs of L and T dwarfs, {\em
  photo-type} can detect unusual L and T dwarfs that may have been
missed in previous searches. Therefore we can check for any bias in
the template colours by looking at the distribution of colours of the
new {\em photo-type} sample. This analysis is presented in
\S\ref{colourcomp}.

We created a multiband $izYJHKW1W2$ photometric catalogue by
  combining SDSS, UKIDSS, and ALLWISE data over 3344\,deg$^2$. The
  starting point was the region of the UKIDSS Large Area Survey Data
  Release 10 (DR10) covered by all four bands, $YJHK$. Point sources,
  in the magnitude range $13.0<J<17.5$, detected in all four bands
  were matched to SDSS DR9 $i$ and $z$ \citep{ahn12}, and ALLWISE $W1$
  and $W2$. One source was undetected in WISE and a handful of sources
  were undetected in the SDSS bands. These undetected sources were
  retained in the catalogue, but the bands in which the source was
  absent were ignored in the fitting. Similarly, sources blended with
  a neighbour in the WISE images were retained, and these bands were
  ignored in the fitting. Since we insist all sources are detected in
  all four bands $YJHK$ we then checked (Paper I, \S3.1) whether this
  meant that sources with extreme colours would be missed (because
  undetected in $Y$, $H$, or $K$).  We undertook a full simulation of
  the colours of each spectral type, using the templates, adding
  appropriate random photometric errors, and measured the proportion
  of sources that fell below the detection limit in any band, over the
  volume of the survey, defined by the sample limit $J=17.5$. The
  result was that total incompleteness of the L and T samples due to
  this effect is substantially less than $1\%$ i.e. brighter than
  $J=17.5$ essentially all L and T dwarfs will be detected in all four
  $YJHK$ bands, so the base sample for the search for L and T dwarfs
  is effectively complete.

Taking a cut at $Y-J>0.8$
left 9487 sources, and classification produced a sample of 1281 L
dwarfs and 80 T dwarfs, which are catalogued in \S\ref{sample}. Of the
190 known L and T dwarfs contained in the parent catalogue of 9487
stellar sources, all 190 (previously 189, see \S\ref{WISEquasars}) are
correctly classified as ultracool dwarfs.

\begin{table}\scriptsize
\centering
\caption{Average photometric errors, by band}
\begin{tabular}{l c c c c c c c c c}
\hline\hline
 & $\sigma_{i}$ & $\sigma_{z}$ & $\sigma_{Y}$ & $\sigma_{J}$ &
$\sigma_{H}$ & $\sigma_{K}$ & $\sigma_{W1}$ & $\sigma_{W2}$ &
$\sigma_{\mathrm All}$ \\  
 Mean   & 0.12 & 0.07 & 0.03 & 0.02 & 0.02 & 0.02 & 0.04 & 0.08 & 0.05
 \\
 Median & 0.09 & 0.06 & 0.03 & 0.02 & 0.02 & 0.02 & 0.04 & 0.07 & 0.03
 \\ \hline 
\end{tabular}
\label{photomerrors}
\end{table}

The accuracy of {\em photo-type} classifications for this sample was
assessed in three ways: i) by comparing the {\em photo-type}
classifications against published spectroscopic classifications, ii)
from our own spectroscopy of sources in the catalogue, and iii) by
creating realistic synthetic catalogues from the template colours and
classifying. We found that {\em photo-type} classifications using all
8 bands are accurate to one subclass r.m.s., or better, at all
magnitudes brighter than $J=17.5$.

The S/N is high for most sources in most bands, which explains
  the accurate classifications. The median photometric error, over all
  bands, for the 1361 L and T dwarfs catalogued, is 0.03 mag., and the
  photometric error is $<0.1$\,mag. for $90\%$ of the photometric
  measurements. In Table 1 we list the mean and
  median photometric error for each band, for the LT sample.

As explained in Paper I (\S4), the uncertainty of $\pm1$ sub-types
  results in a bias \citep{eddington} in the number counts as a
  function of spectral type.\footnote{The bias is, of course, not unique to {\em
      photo-type}. A spectroscopic sample classified with an
    uncertainty of $\pm1$ sub-type would have the same bias.} For
  example, because the counts rise steeply towards earlier types, more
  M9 dwarfs will be scattered into the L0 bin than L0 dwarfs will be
  scattered into the M9 bin. We will correct for this bias in
  computing the luminosity function.

We now describe one minor change made since the completion of Paper I,
and one correction.

\subsection{Extension to include WISE colours for quasars}
\label{WISEquasars}
 
In Paper I quasar template colours were only available for the
$izYJHK$ bands. We have since added the $W1W2$ bands to the quasar
templates, improving the discrimination between peculiar red L and T
dwarfs and reddened quasars. We have reclassified the whole $Y-J>0.8$
catalogue using the improved templates, which resulted in only a very
small number of changes in classification. The total sample size
increased by just five. Significantly, the only previously-known
source that was misclassified, as a reddened quasar, the unusual red L
dwarf \object{2MASS J01262109+1428057} discovered by \citet{metchev08} (see Paper
I, \S3.1), is now correctly classified as an ultracool dwarf. This
source is a very low-gravity young brown dwarf, with spectroscopic
classification L2$\gamma$. So at present there is no evidence that our
sample is incomplete for peculiar red L and T dwarfs. But without a
set of templates for peculiar red L and T dwarfs, which will only
become available once larger samples have been obtained, it is
difficult to quantify accurately the completeness for such sources.

\subsection{Correction: Inaccurate M star templates}

Table 1 in Paper 1 provides our $izYJHKW1W2$ template colours over the
spectral range L0 to T8, as well as an extension to cover the spectral
range M5 to M9. \citet{schmidt15} have pointed out that our $i-z$
template colours for M dwarfs disagree with their $i-z$ colours. We
believe the \citet{schmidt15} $i-z$ colours are correct, and that
  the M5 to M9 template colours in Paper I should not be used. While
differences in the samples used contribute to the discrepancies, the
most significant factor is that we used the Hammer \citep{covey07}
spectral classifications for M stars, straight from the SDSS
database. \citet{west11} showed that for spectral types $>$M5 the
Hammer classifications become systematically offset relative to visual
classifications, in the sense that the visual classifications provide
later spectral types. Starting with M dwarfs visually classified
$>$M5, they found that $38\%$ had Hammer classifications one spectral
sub-type earlier.

In Table \ref{Schmidt Temps} we provide revised template colours for
M7, M8, and M9 dwarfs. We started with the \citet{schmidt15} sample of 11820
M7-M9 dwarfs. From this we produced a matched sample of 3622 dwarfs
(1930 M7, 1060 M8, 425 M9) with
$izYJHKW1W2$ photometry. The tabulated colours are the median colours
for each spectral type. 

We have reclassified all sources using the revised M star template
colours.  Our LT sample is classified to the nearest 0.5 spectral
sub-type. Changing the late M template colours only affects the L0
bin, as it changes the colour boundary between M9.5 and L0. The colour
difference between M9.5 and L0 is now larger in all colours, meaning
that the L0 bin is wider in colour space, so that a significant
fraction of sources previously classified M9.5 should have been
classified L0. Using the revised templates we find that an additional
199 sources are classified L0, significantly increasing the total
sample size to 1361.

\begin{table}\scriptsize
\centering
\caption{Revised template colours for late M dwarfs.}
\begin{tabular}{c c c c c c c c c c c c c c c}
\hline\hline
{SpT} & {$i-z$} & {$z-Y$} & {$Y-J$} & {$J-H$} & {$H-K$} & {$K-W1$} & {$W1-W2$}\\\hline
M7 & 1.36 & 0.55 & 0.68 & 0.54 & 0.38 & 0.17 & 0.20\\
M8 & 1.68 & 0.69 & 0.79 & 0.56 & 0.44 & 0.19 & 0.22\\
M9 & 1.86 & 0.79 & 0.87 & 0.59 & 0.49 & 0.22 & 0.23\\ \hline
\end{tabular}
\tablefoot{The template colours for spectral types M5-M9 in Paper I (Tables 1 and 2) should not be used. All photometry is on the Vega system.}
\label{Schmidt Temps}
\end{table}

\section{Sample of 1361 L and T dwarfs}
\label{sample}

The new sample is presented in Tables \ref{ldwarfs} and \ref{tdwarfs},
listing the coordinates, the 8-band photometry, the {\em photo-type}
classification, and the $\chi^2$ of the fit, for the 1281 sources
classified as L dwarfs and the 80 sources classified as T dwarfs,
respectively. Sources have been classified to the nearest half
sub-type, by interpolating the template colours (Table \ref{Schmidt
  Temps} in this paper, and Table 1 of Paper I). Also listed are any
existing spectroscopic classifications, and the relevant
reference. The large majority of the {\em photo-type} classifications
are based on photometry in all 8 bands. Sources without $W1$ and $W2$
photometry, primarily due to blending, are marked e.g. L2:, indicating
that the classification is less certain. The same is the case for
sources undetected in both $i$ and $z$. The classifications for the
handful of sources with only $YJHK$ photometry are marked
e.g. T4::. All sources have been inspected in the images in all
bands. Nevertheless we recommend scrutiny of the images prior to any
spectroscopic observations, particularly for peculiar sources.

\begin{sidewaystable}\scriptsize
\caption{Sample of 1281 L dwarfs}
\begin{tabular}{l l l l l l l l l l l l l l l l l l l l l}
\hline\hline
Name & {i} & {ierr} & {z} & {zerr} & {Y} & {Yerr} & {J} & {Jerr} & {H} & {Herr} & {K} & {Kerr} & {W1} & {W1err} & {W2} & {W2err} & {PhT} & {$\chi^2$} & {SpT} & {Ref}\\\hline  
ULAS J000005.87+152354.4 & 21.30 & 0.13 & 19.50 & 0.12 & 18.47 & 0.04 & 17.27 & 0.03 & 16.51 & 0.03 & 15.92 & 0.03 & 99.00 & 99.00 & 99.00 & 99.00 & L2: & 2.22 & 99 & 99 \\
ULAS J000100.45+065259.6 & 18.62 & 0.02 & 16.56 & 0.02 & 15.70 & 0.01 & 14.76 & 0.01 & 14.08 & 0.01 & 13.54 & 0.01 & 13.33 & 0.02 & 13.02 & 0.03 & L0 & 5.62 & 99 & 99 \\
ULAS J000112.24+153534.3 & 19.92 & 0.04 & 17.99 & 0.03 & 16.88 & 0.01 & 15.46 & 0.01 & 14.48 & 0.01 & 13.62 & 0.01 & 12.97 & 0.02 & 12.54 & 0.02 & L5.5p & 55.82 & L4 & 11 \\
\end{tabular}
\tablefoot{Only the first three lines of the table are provided. The full table is available at the CDS. References: (1) \citealt{schmidt07}, (2) \citealp{reid08}, (3) \citealp{west11}, (4) \citealp{west08}, (5) \citealp{hawley02}, (6) \citealp{zhang10}, (7) \citealp{schmidt10}, (8) \citealp{schneider02}, (9) \citealp{kirkpatrick11}, (10) \citealp{scholz09}, (11) \citealp{knapp04}, (12) \citealp{smith14}, (13) \citealp{aberasturi11}, (14) \citealp{testi09}, (15) \citealp{faherty09}, (16) \citealp{kirkpatrick10}, (17) \citealp{fan00}, (18) \citealp{allen07}, (19) \citealp{metchev08}, (20) \citealp{liu06}, (21) \citealp{bihain10}, (22) \citealp{berger06}, (23) \citealp{chiu06}, (24) \citealp{cushing06}, (25) \citealp{leggett07}, (26) \citealp{kirkpatrick05}, (27) \citealp{marocco13}, (28) \citealp{skrzypek15}, (29) \citealp{DayJones13}}
\label{ldwarfs}
\end{sidewaystable}

\begin{sidewaystable}\scriptsize
\caption{Sample of 80 T dwarfs}
\begin{tabular}{l l l l l l l l l l l l l l l l l l l l l}
\hline\hline
Name & {i} & {ierr} & {z} & {zerr} & {Y} & {Yerr} & {J} & {Jerr} & {H} & {Herr} & {K} & {Kerr} & {W1} & {W1err} & {W2} & {W2err} & {PhT} & {$\chi^2$} & {SpT} & {Ref}\\\hline  
ULAS J000844.34+012729.4 & 99.00 & 99.00 & 99.00 & 99.00 & 18.20 & 0.04 & 16.99 & 0.02 & 17.40 & 0.06 & 17.54 & 0.10 & 17.02 & 0.14 & 14.83 & 0.07 & T6.5: & 1.26 & 99 & 99 \\
ULAS J003451.98+052306.8 & 24.09 & 0.61 & 18.38 & 0.04 & 16.21 & 0.01 & 15.14 & 0.01 & 15.58 & 0.01 & 16.07 & 0.03 & 15.09 & 0.04 & 12.55 & 0.03 & T7p & 45.15 & T6.5 & 1 \\
ULAS J004730.55+113222.5 & 22.99 & 0.42 & 19.92 & 0.17 & 18.37 & 0.04 & 17.17 & 0.03 & 16.77 & 0.04 & 16.90 & 0.06 & 16.30 & 0.07 & 15.11 & 0.09 & T3 & 15.59 & 99 & 99 \\
\end{tabular}
\tablefoot{Only the first three lines of the table are provided. The full table is available at the CDS. References: (1) \citealp{faherty09}, (2) \citealp{metchev08}, (3) \citealp{liu06}, (4) \citealp{chiu06}, (5) \citealp{burgasser06b}, (6) \citealp{scholz12}, (7) \citealp{burningham10}, (8) \citealp{gelino14}, (9) \citealp{burgasser06a}, (10) \citealp{pinfield08}, (11) \citealp{hawley02}, (12) \citealp{mace13}, (13) \citealp{burgasser04b}, (14) \citealp{burningham11}, (15) \citealp{scholz10}, (16) \citealp{burningham10b}, (17) \citealp{kirkpatrick11}, (18) \citealp{smith14}, (19) \citealp{bardalez14}, (20) \citealp{burningham13}, (21) \citealp{skrzypek15}}
\label{tdwarfs}
\end{sidewaystable}

\begin{table}
\caption{Number counts by spectral type}
\begin{tabular}{l r | l r}
\hline\hline
{SpT} & {Count} & {SpT} & {Count} \\\hline
L0 & 596 & T0 & 10 \\
L1 & 279 & T1 & 11 \\
L2 & 163 & T2 & 7\\
L3 & 92 & T3 & 8\\
L4 & 75 & T4 & 13\\
L5 & 32 & T5 & 14\\
L6 & 10 & T6 & 5\\
L7 & 18 & T7 & 11\\
L8 & 8 &T8 & 1 \\
\cline{3-4}
L9 & 8 & & \\
\cline{1-2}
\end{tabular}
\label{counts}
\tablefoot{Here each bin is a full spectral sub-type e.g. L4 and L4.5 have been combined into the L4 bin.}
\end{table}

\begin{figure}
\centering
\includegraphics[width = 9.cm]{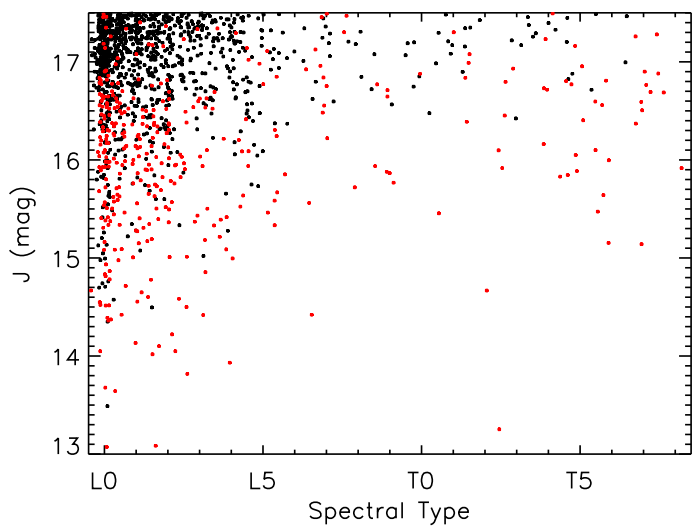}
\caption{Plot of $J$ mag. against spectral type for the 1361 L and T
  dwarfs in the {\em photo-type} sample. Red symbols indicate
  previously catalogued sources, while black symbols are new
  discoveries. The spectral types have been determined to the nearest
  half sub type, but small random offsets have been added for this plot to
  separate overlapping points.}
\label{specj}
\end{figure}

\begin{figure}
\centering
\includegraphics[width = 9.cm]{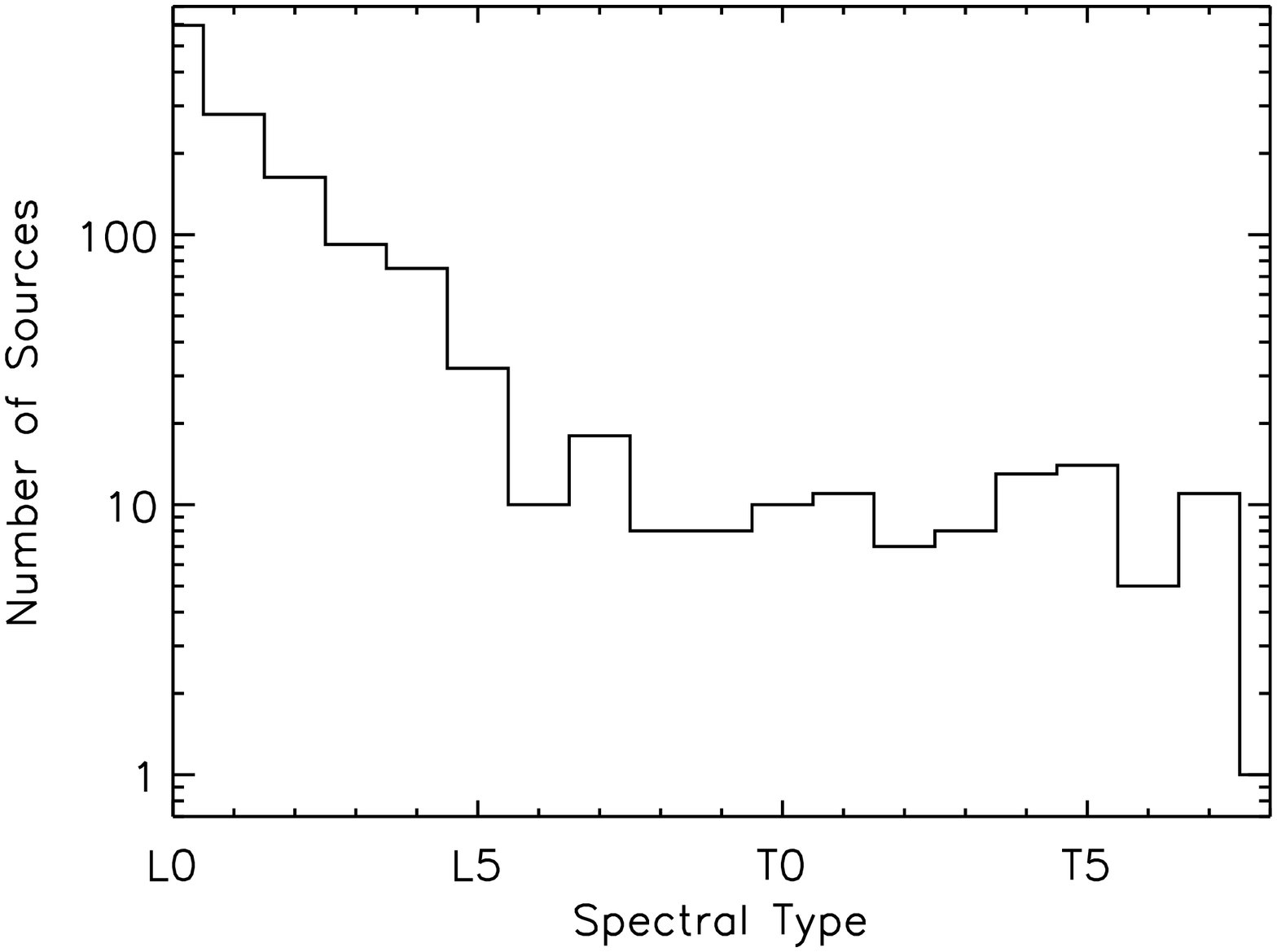}
\caption{Histogram of spectral type for the sample of 1281 L dwarfs
  and 80 T dwarfs. Here each bin is a full spectral sub-type e.g. L4 and L4.5 have been combined into the L4 bin.}
\label{spechist}
\end{figure}

\begin{figure}
\centering
\includegraphics[width = 9.cm]{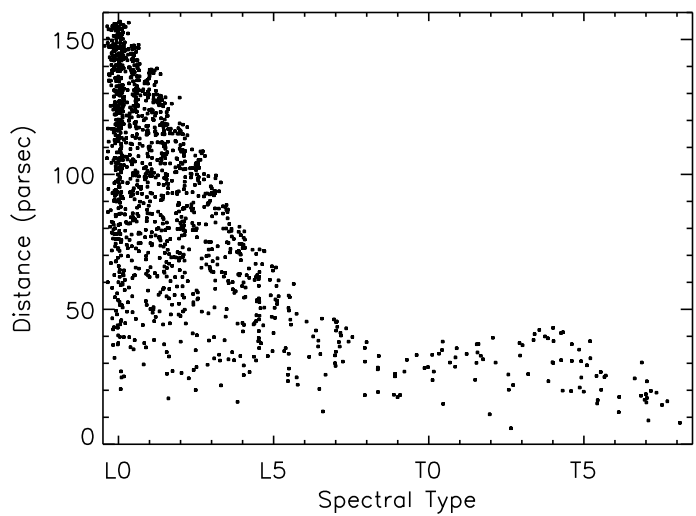}
\caption{Plot of distance against spectral type for the 1361 L and T
  dwarfs in the {\em photo-type} sample.  The distances have been
  estimated from the $J$ magnitude, using the relation between $M_J$
  and spectral type of \citet{dupuy12}. The spectral types have been
  determined to the nearest half sub type, but small random offsets
  have been added for this plot to separate overlapping points.}
\label{SpT_d}
\end{figure}

\begin{figure}
\centering
\includegraphics[width = 9.cm]{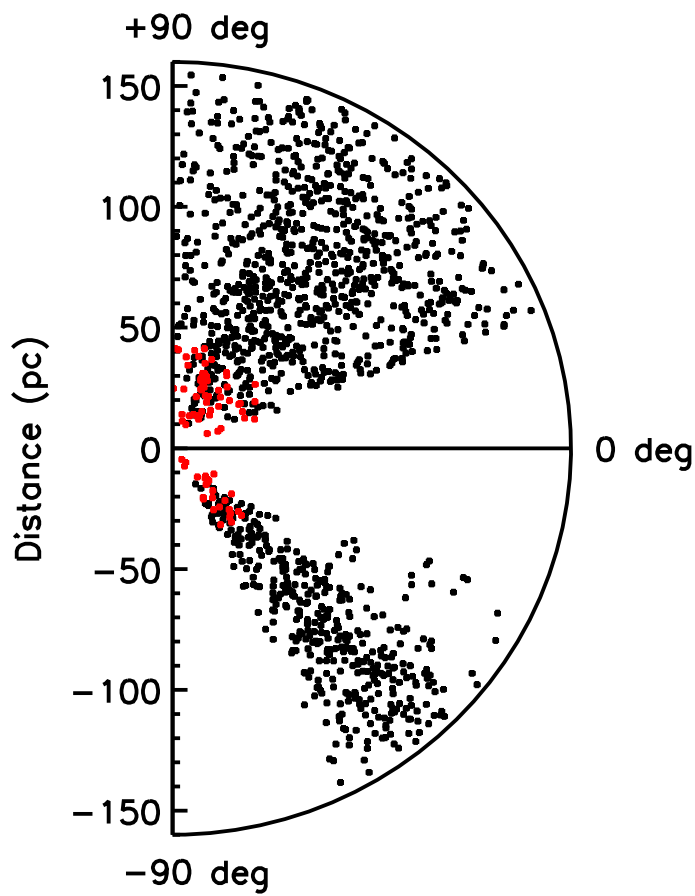}
\caption{Pot of distance against Galactic latitude, in polar
  coordinates, for the 1281 L dwarfs (black) and 80 T dwarfs (red). }
\label{galactic_dis}
\end{figure}

\begin{figure}
\centering
\includegraphics[width = 9.cm]{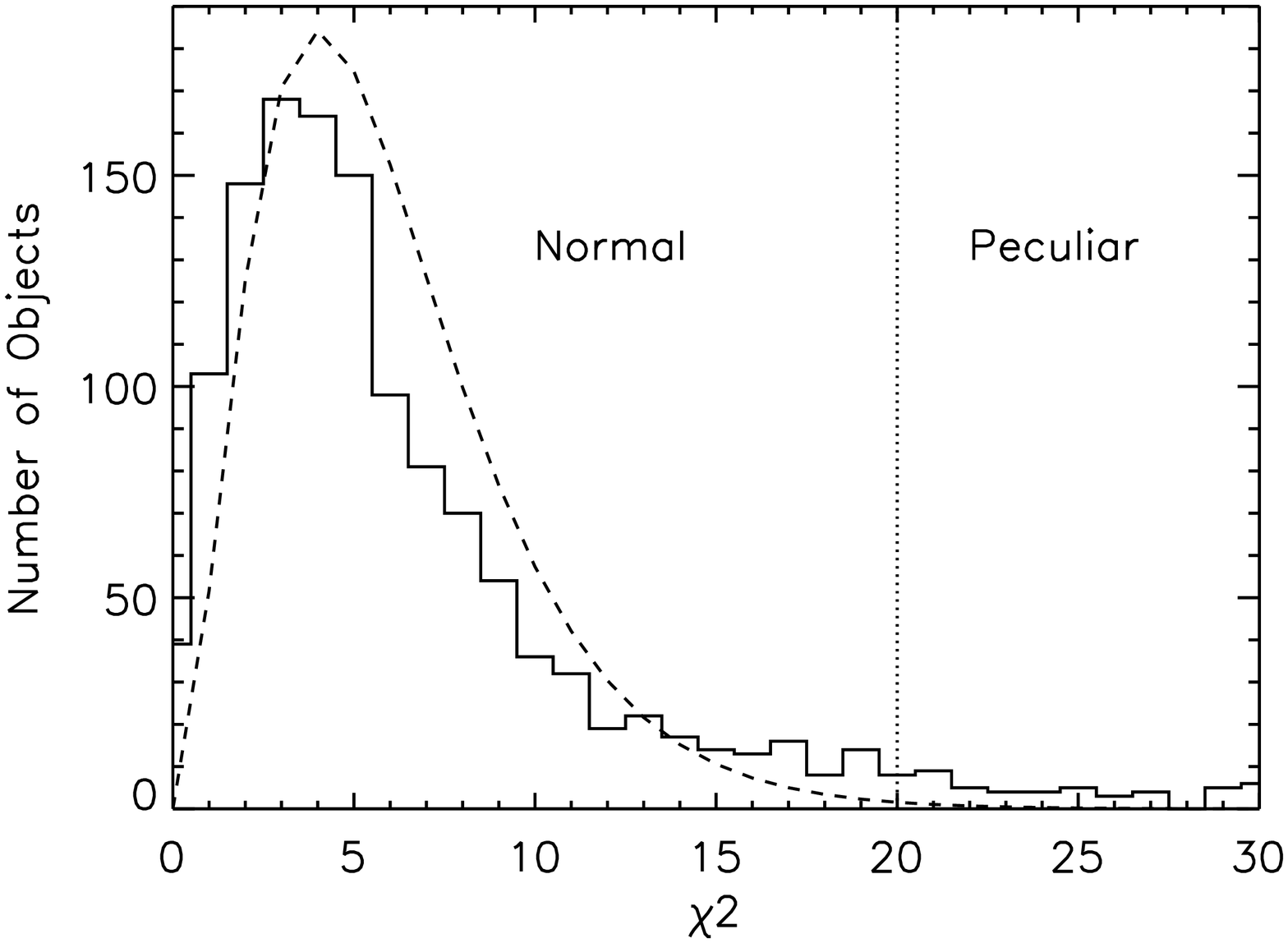}
\caption{Histogram of the distribution of $\chi^2$ for the 1361 L and
  T dwarfs, compared to the theoretical distribution for $\nu=6$
  degrees of freedom, plotted as the smooth curve.}
\label{chisq}
\end{figure}

The general properties of the sample are illustrated in a set of
plots, Figs \ref{specj} to \ref{colour3}. Fig. \ref{specj} plots the
$J$ mag. against {\em photo-type} spectral type. It shows that
previous samples are fairly complete to $J=16$, with 154 of the
  210 sources previously catalogued, i.e. $73\%$, with incompleteness increasing
progressively towards fainter magnitudes. In total 998 of the 1361
sources are new. Fig. \ref{spechist} plots the distribution of
spectral types as a histogram. The number counts for this plot are
provided in Table \ref{counts}. The steep decline in number counts
from L0 to L7 and subsequent flattening across the T types is a
reflection of the volume surveyed, as illustrated in Fig. \ref{SpT_d},
which plots distance against spectral type for the sample.  We used
the relation between the absolute magnitude in the $J$ band, $M_J$,
and spectral type from \citet{dupuy12} to estimate distances.  Dwarfs
of spectral type L0 are detectable out to 150\,pc, but the limiting
distance drops rapidly towards later types, and then flattens off near
L6. Over most of the T sequence the distance limit is in the range
30-40\,pc. It is evident from the plot that the space density as a
function of spectral type does not vary strongly over most of the
  spectral range.  In Fig. \ref{galactic_dis} we plot distance
against Galactic latitude $b$, in polar coordinates, with L dwarfs
plotted black, and T dwarfs plotted red. This illustrates the fact
that the LAS fields lie mostly at high Galactic latitudes. The
variation in numbers with $b$ reflects the variation in the solid
angle surveyed with $b$. Although there is a measureable decline in
space density with distance from the Galactic plane, the sample
reaches insufficient depth to be useful for constraining the scale
height of the L and T populations on its own, but will be very
  useful when supplemented with a deep sample.

In Fig. \ref{chisq} we plot the histogram of $\chi^2$ values for the
sample, compared against the theoretical curve for $\nu=6$ degrees of
freedom. For most of the sample we have photometry in 8 bands. The
brightness of the source is a free parameter in fitting templates, and
the spectral type is treated as a second free parameter. The actual
$\chi^2$ distribution is different to the theoretical curve, and has a
pronounced tail.  This means that our photometric model, where we
added an error $\Delta m=0.05$\,mag. in each band to account for the
spread in colours, does not fully model the variation in the
population, perhaps due to correlations between bands for peculiar
sources. There are 97 sources with $\chi^2>20$, i.e. $7\%$ of the
sample, and we have used this limiting value to define a sample of
peculiar objects. These sources are marked e.g. L3p in the
catalogue. The proportion of T dwarfs classed peculiar, 22/80,
compared to 75/1281 for L dwarfs, is disproportionately high, implying
that the scatter in colours is larger for T dwarfs than for L
dwarfs. Using the criterion $\chi^2>35$ for T dwarfs reduces the
proportion to 7/80. Spectroscopy of 9 objects with $\chi^2>20$ is
presented in \S\ref{spectra}.


\subsection{Colour relations for the {\em photo-type} sample}
\label{colourcomp}

We now return to the question raised in Paper I of whether the colours
of the 190 L and T dwarfs used in deriving the colour polynomials, on
which the {\em photo-type} method rests, are representative of the
full L and T population. As noted in Paper I, because the SEDs of the
potential contaminating populations, M stars and reddened quasars, are
so different to L and T dwarfs, the {\em photo-type} method can
potentially identify L and T dwarfs that are quite different to
typical L and T dwarfs, that might have been missed by previous
searches. These would manifest themselves as a cloud of sources with
colours significantly different from the template colours.

\begin{figure*} 
\includegraphics[width=16.5cm]{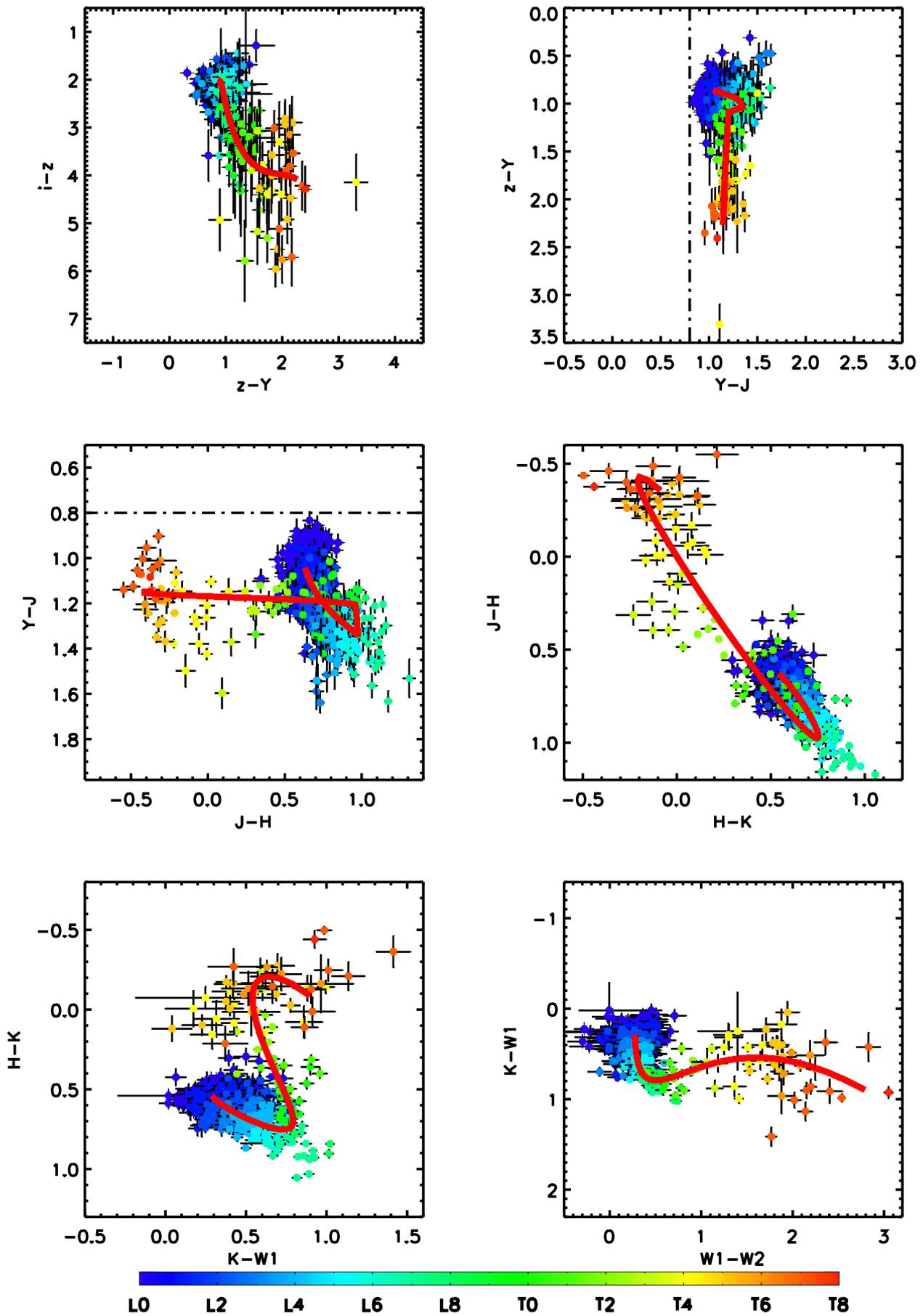} 
\caption{Two colour diagrams for the new sample of 1361 L and T
  dwarfs, colour coded by spectral type according to the
  colourbar. The colour cut $Y-J>0.8$ is marked.} 
\label{rainbow}
\end{figure*}

Fig. \ref{rainbow} plots two-colour diagrams, successively cycling
through pairs of colours from the set $i-z$, $z-Y$, $Y-J$, $J-H$,
$H-K$, $K-W1$, $W1-W2$. In each plot the 1361 objects are represented
by rainbow colours from blue to red, that translate to classifications
L0 to T8, as shown by the colourbar. The red line in each panel plots
the template colour relations. We note the following points:
\begin{enumerate}

\item The scatter in the diagrams is larger than can be explained by
  the photometric errors. In Paper I, in classifying, we allowed for
  this by adding (in quadrature) an uncertainty to each point in each band of 0.05\,mag.,
  corresponding to an uncertainty of 0.07\,mag. in each colour.

\item As noted in Paper I, the $i-z$ template colours for T dwarfs are
  not well defined, and the $i$ band does not contribute usefully to
  the classification of T dwarfs.

\item There are very few objects with colours close to the colour cut
  $Y-J=0.8$, meaning there is no evidence we have missed a significant
  number of sources due to this colour cut. \footnote{In fact we
    checked explicitly that there are no sources in the colour range
    $0.7<Y-J<0.8$ classified L or T.}

\item There is a suggestion that mid-T dwarfs are mostly redder in
  $Y-J$ than the template curve (sources near $Y-J=1.3$, $J-H=0.0$).

\item There is also evidence for a mismatch between the colours and
  the template curve in the $J-H$ vs $H-K$ plot, near $J-H=0.3$, where
  T3 dwarfs lie bluer in $H-K$ than the curve.

\item Referring to the $H-K$ vs $K-W1$ diagram, there are several mid
  T dwarfs, around $H-K\sim 0$ that have blue $K-W1$ colours compared
  to the red curve. This suggests that the template polynomial
  (Fig. 2, Paper I) should bend to bluer colours near T4. Nevertheless
  making this correction would have very little effect on the
  classifications, which around T4 are largely determined by the $z-Y$
  and $W1-W2$ colours.

\end{enumerate}

Variability may contribute to the scatter in these plots, as not all
bands were observed at the same epoch.  In UKIDSS DR10, $8\%$ of the
area has $J$ observations at two epochs. For this work we have always
used the first epoch $J$ observation, which may not be the nearest in
time to the $YHK$ observations. In looking at sources in the catalogue
with high values of $\chi^2$, the possibility that variability may
contribute to the poor fit should be considered, and a check against
the observation dates made.

\begin{figure}
\centering
\includegraphics[width = 17.5cm, trim=5.cm 6.cm 0cm 4.5cm, clip=true]{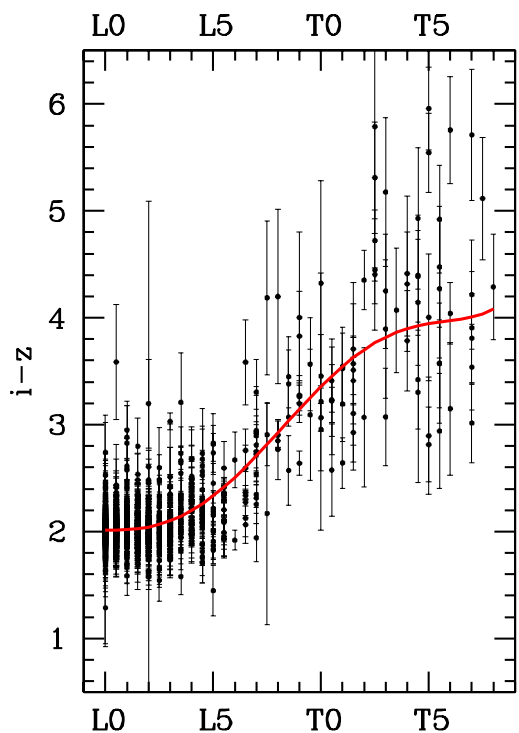}
\caption{Plot of $i-z$ colour vs.  {\em photo-type} spectral sub-type
  for LT dwarfs in the {\em photo-type} sample. The red curve plots
  the template colours from Paper I. All photometry is on the
    Vega system.}
\label{colour1}
\end{figure}

\begin{figure}
\centering
\includegraphics[width = 17.5cm, trim=5.cm 0cm 0cm 0cm, clip=true]{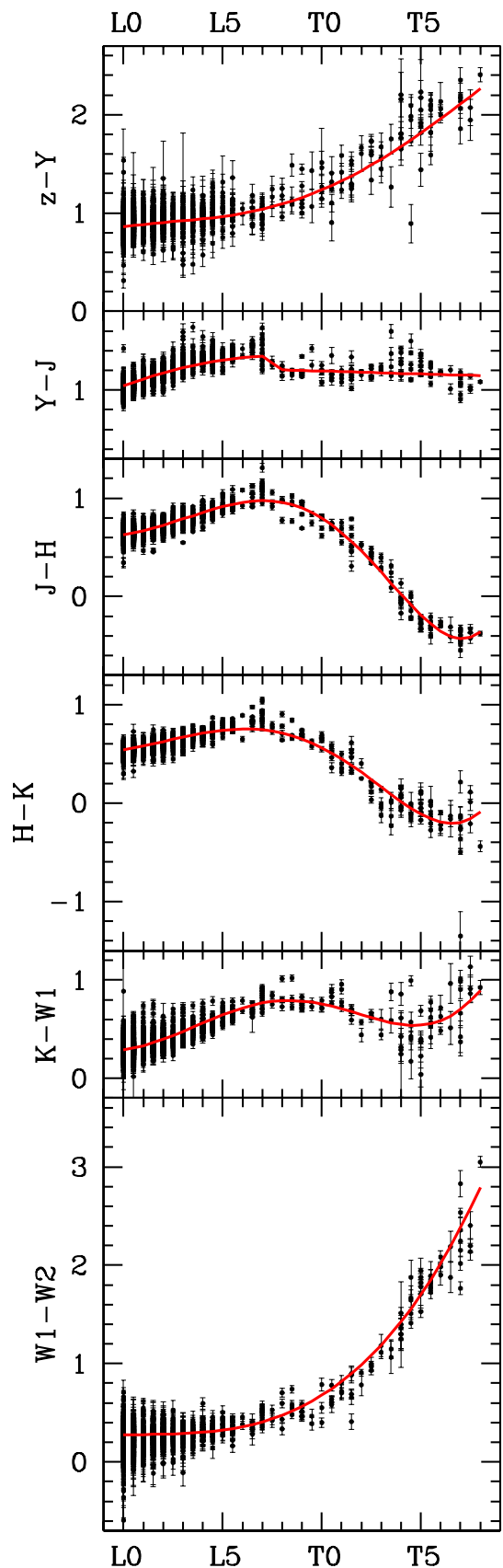}
\caption{Plot of colours $z-Y$, $Y-J$, $J-H$, $H-K$, $K-W1$, $W1-W2$
  vs.  {\em photo-type} spectral sub-type for LT dwarfs in the {\em
    photo-type} sample. In each panel the red curve plots the template
  colours from Paper I. All photometry is on the Vega system.}
\label{colour2}
\end{figure}

\begin{figure}
\centering
\includegraphics[width = 17.5cm, trim=5.cm 6.cm 0cm 4.5cm, clip=true]{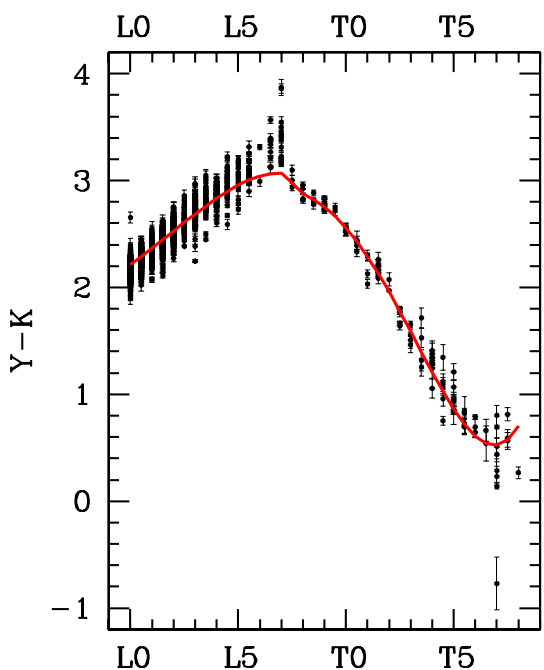}
\caption{Plot of $Y-K$ colour vs. {\em photo-type} spectral sub-type
  for LT dwarfs in the {\em photo-type} sample. The red curve plots
  the template colours from Paper I. All photometry is on the
    Vega system.}
\label{colour3}
\end{figure}

The features noted in the two-colour diagrams, listed above, may also
be picked out in Figs \ref{colour1} and \ref{colour2}, that plot
colour against {\em photo-type} spectral type, together with the
template polynomials. None of these features is sufficiently striking
to suggest that the templates need changing at this time, but they
motivate spectroscopic investigation of some of the outliers. There is
nevertheless one additional feature that suggests the presence of a
population of objects that was under-represented in the original
sample of 190 known sources used in creating the templates. This is a
group of very red objects evident around spectral type L7, where the
template curves underfit the colours in the $Y-J$, $J-H$, and $H-K$
plots. This feature is accentuated in Fig \ref{colour3}, where we plot
$Y-K$ against spectral type. In this plot a dramatic discontinuity in
colour $\Delta(Y-K)\sim0.4$\,mag. is evident, between spectral types
L7 and L8.

The explanation for this discontinuity is not clear, but 
three separate effects may contribute. First, it is possible that the actual curvature of the $J-H$ and $H-K$
colour relations around L7 is inadequately represented by the low-order polynomials used.
Second, there are several objects that are very
red in $Y-K$, that may not be genuine L7s but are classified as such
because, over the near-infrared bands, this is the reddest spectral
type. An example is the L2$\gamma$ dwarf \object{2MASS J01262109+1428057}
\citep{metchev08} previously discussed. The {\em photo-type}
classification of this source is L7p, and it is one of two objects
with $Y-K\sim3.9$. These very red objects may make the discontinuity
appear larger than it really is i.e. they should really be outliers
plotted at a different spectral type. Third, there is a discontinuity
in the $Y-J$ template curve of size 0.14 mag. between types L7 and
L8. In Paper I we speculated that this was associated with a rapid
weakening of FeH absorption in the $Y$ band. We would expect
photometric errors to tend to wash out this feature in the
classification process, yet in Figs \ref{colour2} and \ref{colour3}
the feature appears to be enhanced relative to the plot in Paper
I. Therefore the discontinuity appears to be real, and requires
explanation. Near-infrared spectroscopy of several sources in the
catalogue in the interval L6 to L9 could be very revealing.

\section{Sample completeness}
\label{completeness}

In this section we quantify the completeness of the sample. In Paper
I, \S2.2.1, we showed that the SEDs of quasars and L and T dwarfs are
sufficiently distinct that contamination of the LT sample by reddened
quasars should be negligible. This also means that any L or T dwarf in
the base sample of 9487 stars, $Y-J>0.8$, should be correctly
classified as such, modulo an uncertainty in classification of one
spectral sub-type (meaning that some Ls are classified M and {\em vice
  versa}). This conclusion rests on the assumption that the L and T
templates and the quasar templates are adequate representations of the
colours of these populations. As noted in \S\ref{WISEquasars} and
\S\ref{colourcomp}, the polynomial modelling of the colours of the reddest L
dwarfs is not entirely satisfactory. Although at present there is no
evidence that any such sources are missed by the {\em photo-type}
method, until the modelling of very red sources is improved it is not
possible to be definitive on this matter.

As described in Paper I we searched DwarfArchives.org and several
recent papers for L and T dwarfs $13.0<J<17.5$ within the survey
footprint. Here we use these objects to identify potential sources of
incompleteness that are not addressed by the colour modelling
presented in Paper I. There are three close binaries, classified as
stellar (i.e. a point source) in 2MASS, but as non-stellar\footnote{As
  described in Paper I, we used the UKIDSS parameter {\tt
    mergedclassstat} to define classes stellar and non-stellar} in
UKIDSS, because of the better image quality, and therefore
missed. There is therefore a small bias against finding binaries with
separations of a few tenths of an arcsec. Rather than attempt to
quantify this, we simply define our sample as consisting of objects
classified as stellar in UKIDSS. The remainder of the sample is used
for identifying the different sources of incompleteness, which are as
follows.

\begin{enumerate}

\item A handful of objects are missed because of unreliable photometry
  in any band e.g. landing on a bad row in one of the SDSS
  images. This left 192 known L and T dwarfs $J<17.5$ with good photometry
  that could have been selected by {\em photo-type}.

\item One of the 192 sources, \object{WISEPC J092906.77+040957.9}, was
  missed due to large proper motion, that just exceeded the UKIDSS
  $YJHK$ inter-band $2\arcsec$ matching radius. No sources were missed
  from large proper motion when matching to SDSS and WISE for which a
  larger $10\arcsec$ match radius was used.

\item Another of the 192 sources, \object{SDSS J074656.83+251019.0}, was missed
  because it has $Y-J<0.8$ i.e. it is a peculiar blue source. The
  {\em photo-type} classification of this source is M8.5.

\end{enumerate}

Of the remaining 190 known L and T dwarfs, all were successfully
classified as ultracool dwarfs, so the {\em photo-type} method {\em
  per se} does indeed appear to be highly complete. In the following
three subsections \S\S4.1, 4.2, 4.3, we quantify the incompleteness
associated with points 1, 2, 3, above, respectively.

\subsection{Unreliable photometry}

In creating a clean multi-band catalogue of matched photometry of
stellar sources, a proportion of sources will be eliminated due to
unreliable photometry. There are several causes of unreliable
photometry. For example \citet{dye06} note various artifacts in the
UKIDSS data including moon ghosts, channel bias offsets, and
cross-talk. Then in the SDSS data a clean UKIDSS source may land on a
diffraction spike or bad row. Alternatively two separate sources in
UKIDSS (seeing FWHM $\sim 0.8\arcsec$) may be blended in the SDSS images
(seeing FWHM $\sim 1.4\arcsec$). The different sources of incompleteness
may affect only one band (e.g. bad row), or more than one
(e.g. diffraction spike). It is difficult to quantify all these
sources one by one, but in matching several bands together the total
fraction lost can be significant.

We have made an empirical estimate of the fraction of sources lost due
to the cumulative effects of unreliable photometry in different bands
by using deeper UKIDSS Deep Extragalactic Survey (DXS) data as a
reference. The DXS \citep{UKIDSS} covered tens of square degrees in
$JHK$ to depths some 3 mag. deeper than the LAS. Our starting point is
to assume that a catalogue of stellar sources $J<17.5$ detected in the
$J$ band in the DXS is a close approximation to a complete sample of
isolated stellar objects in the field. The DXS $J$ image is formed
from a stack of many images. In averaging to form the stack,
discrepant images are eliminated, meaning the DXS $J$ catalogue should
be very clean. The DXS overlaps the LAS in the SA22 field (centre
$22^h$ $17^m$, $+00^\circ$ $20^\prime$), and we selected a catalogue
of stellar sources over 5.9 deg$^2$. We then measured how many of
these propagated through to the base catalogue of LAS sources
$13<J<17.5$ in $izYJHK$ that was the starting point for our search for
L and T dwarfs. Note that because of the problem of blending we did
not require a successful match to W1 and W2 to include an object in
the catalogue (see Paper I for more details), so the matching to WISE
is not relevant to the calculation of incompleteness. The result of
the match to DXS was that $8.2\%$ of sources are lost due to
unreliable photometry in one or more bands. The incompleteness is
independent of brightness. Because of this, we account for this source of
incompleteness by a reduction in the effective area of the survey from
3344 to 3070 deg$^2$.

\subsection{Large proper motion}

The search radii for matching to SDSS and WISE were sufficiently large
to capture all known L and T dwarfs. However a few objects with high
proper motion, such as \object{WISEPC J092906.77+040957.9}, could be missed because
of the smaller match radius used in UKIDSS. For $YJHK$ detected
sources in the UKIDSS LAS, the $Y$ band position is used as the
reference point for a $2.0\arcsec$ matching radius i.e. if the
measured offset in $J$, $H$, or $K$ is larger than $2.0\arcsec$, the
source remains unmatched, and it is declared a separate
source. Therefore whether a source is matched depends on the proper
motion, and the $YJ$, $YH$, and $YK$ epoch differences.

\begin{figure} 
\includegraphics[width=9.0cm]{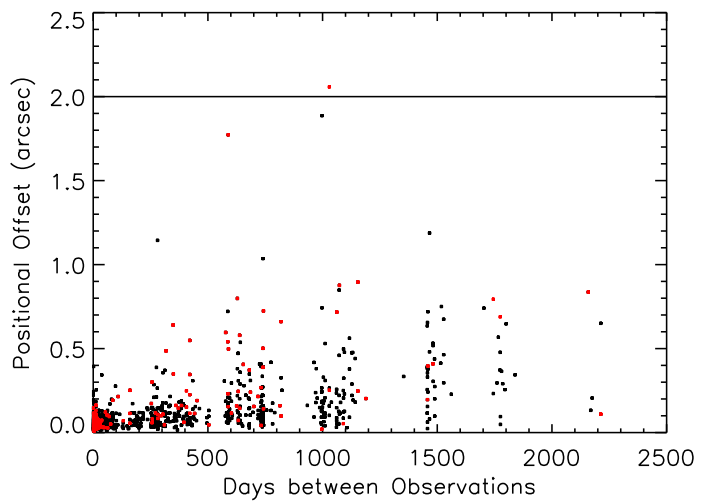} 
\caption{Check of incompleteness due to proper motion in the UKIDSS
  dataset. The positional offsets in the UKIDSS $J$, $H$, and $K$
  images, relative to the reference $Y$ image, were first computed, as
  well as the epoch differences between the pairs of observations. The
  maximum offset was selected and this quantity is plotted for each
  source against the relevant epoch difference. The red dots represent
  the 192 known L and T dwarfs in DwarfArchives.org, $J<17.5$, within
  the UKIDSS footprint. The black dots are the objects in our new
  sample. On the basis of this plot, incompleteness due to proper
  motion is estimated to be $\ll 1\%$.}
\label{properm} 
\end{figure}

To assess the importance of proper motions we took the 192 known L and
T dwarfs and extracted the largest angular offset of the three values
i.e. from the $YJ$, $YH$, and $YK$ matches. These maximum offsets are
plotted in Fig. \ref{properm} against epoch difference, as red
symbols. The source \object{WISEPC J092906.77+040957.9} is the dot plotted above the
line, which marks the $2.0\arcsec$ matching radius. Only one other
catalogued source, the sdL7 \object{2MASS J11582077+0435014} (from
\citealt{kirkpatrick10}), has an offset larger than
$1.0\arcsec$. Given that the catalogued sources are mostly relatively
bright compared to our magnitude limit, their typical proper motions
are likely to be larger than for our sample as a whole, because the
sources are mostly nearer. For example the median distances of the
sources in the sample of 427 late-type M, L, and T dwarfs of
\citet{faherty09} are 23, 29, and 15 pc respectively, whereas the
median distance for our sample as a whole is 94 pc.  So we can expect
incompleteness due to proper motion in our sample to be $<1\%$. The
black symbols show the offsets for the sample of 1361 L and T dwarfs
presented here. With only a handful of sources with offsets
$>1\arcsec$, this is supporting evidence that incompleteness due to
proper motion is very small i.e. there is no evidence for a
significant population of sources with large proper motions.

%\end{itemize}

\begin{figure} 
\includegraphics[width=9.0cm]{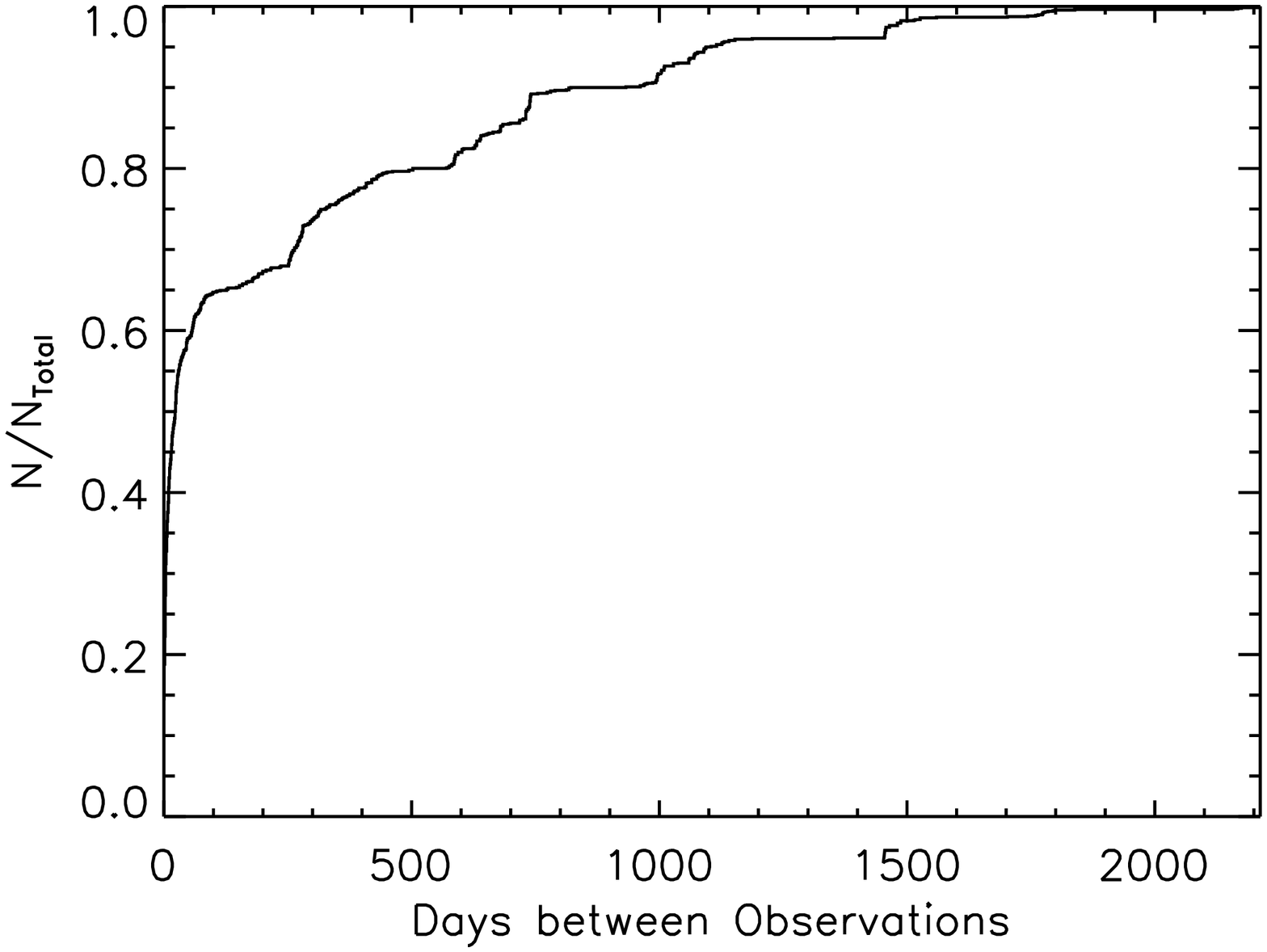} 
\caption{Proportion of fields for which all $YJHK$ observations
  were completed within the time interval.}
\label{obs_time} 
\end{figure}

The reasons why incompleteness due to proper motion is such a minor
issue for the new sample are that most of the new sources are more
distant than the majority of known L and T dwarfs, and because for
most of the fields all of the $YJHK$ UKIDSS data were taken over a
short time period; for over $85\%$ of the fields all filters were
observed within two years of each other. The latter point is
illustrated in Figure \ref{obs_time}, which plots the proportion of
fields for which all the observations were completed within the given
time interval.

In this subsection we quantify incompleteness of the sample in a
physical way, as a function of tangential velocity. We first assume,
for simplicity, that the distribution of maximum time intervals for
the sample $f(t)$ (the black dots in Fig. \ref{properm}) is an
adequate approximation of the distribution of maximum time intervals
over all the UKIDSS fields in which L and T dwarfs could be
found. Next we estimate the distances for the sample using the $J$
magnitudes and the absolute magnitudes from \citet{dupuy12}. We assume
this distribution of distances $g(d)$ is an adequate representation of
the true distribution of the distances of L and T dwarfs in the
field. This will be reasonable provided incompleteness overall is
small, an assumption we can check at the end. We now treat these two
distributions, $f(t)$ and $g(d)$, as independent. It is then possible
to quantify completeness in terms of tangential velocity $v_t$. We
imagine all sources have a particular tangential velocity $v_t$. For
sources at distance $d$ we compute the proper motion, and compute the
fraction that move more than $2.0\arcsec$, from the distribution
$f(t)$. Integrating over $g(d)$ gives the total fractional
incompleteness for the given $v_t$.

\begin{figure} 
\includegraphics[width=9.0cm]{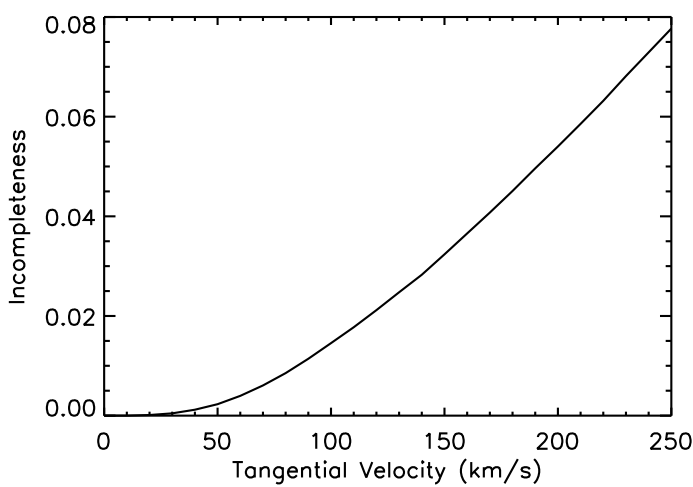} 
\caption{Incompleteness due to proper motion exceeding the UKIDSS
  match radius, as a function of tangential velocity.}
\label{vtan} 
\end{figure}

The results of this calculation are shown in Fig. \ref{vtan}, which
plots the incompleteness as a function of $v_t$. From this plot it is
clear that for L and T dwarfs in the Galactic thin disk, for which the
3D velocity dispersion is $\sigma\sim50$km\,s$^{-1}$ \citep{seifahrt10},
incompleteness due to proper motion is completely negligible. For the
rarer members of the thick disk or halo, incompleteness due to proper
motion is still very small. For a tangential velocity of
$v_t=100\,(200)$\,km\,s${^{-1}}$, the sample incompleteness is only
1.5\,(5.5)$\%$. But such sources are rare. In the sample of 427 late-type M,
L, and T dwarfs analysed by \citet{faherty09}, only 14, i.e. $3\%$,
had tangential velocities greater than 100\,km\,s${^{-1}}$. The overall
incompleteness due to proper motion is therefore negligibly
small, $\ll 1\%$. This conclusion justifies the original assumption that $g(d)$
is an adequate representation of the distribution of distances of the
LT population. Note that the incompleteness values computed are for
the sample as a whole, and that incompleteness depends on magnitude
i.e. the percentage incompleteness is larger for brighter (nearer)
sources.

\begin{table*}\scriptsize
\caption{Coordinates, photometry and classifications of known
L subdwarfs.}
\begin{tabular}{c c c c c c c c c c l l r l}
\hline\hline
{R.A. (2000)} & {Decl. (2000)}& {$i\pm \sigma_i$}&{$z\pm
\sigma_z$}&{$Y\pm \sigma_Y$}&{$J\pm \sigma_J$}&{$H\pm
\sigma_H$}&{$K\pm \sigma_K$}&{$W1\pm \sigma_{W1}$}&{$W2\pm
\sigma_{W2}$}& {SpT} & {PhT} & {$\chi^2$} \\ \hline
02:12:58.07 & +06:41:17.6 & 20.76$\pm$0.08 & 18.87$\pm$0.08 & 18.20$\pm$0.03 & 17.43$\pm$0.03 & 17.06$\pm$0.03 & 16.78$\pm$0.05 & 16.31$\pm$0.06 & 15.98$\pm$0.18 & sdL0.5$^1$ & M6.5 & 19.81\\
03:33:50.84 & +00:14:06.1 & 18.85$\pm$0.02 & 17.27$\pm$0.03 & 16.81$\pm$0.01 & 16.11$\pm$0.01 & 15.77$\pm$0.01 & 15.50$\pm$0.02 & $\--$ & $\--$ & sdL0$^2$ & M7p & 35.19\\
11:58:20.77 & +04:35:01.4 & 20.65$\pm$0.08 & 17.62$\pm$0.03 & 16.61$\pm$0.01 & 15.43$\pm$0.01 & 14.88$\pm$0.01 & 14.37$\pm$0.01 & 13.70$\pm$0.03 & 13.36$\pm$0.03 & sdL7$^3$ & L3p & 128.69\\
12:44:25.90 & +10:24:41.9 & 19.12$\pm$0.02 & 17.47$\pm$0.02 & 16.98$\pm$0.01 & 16.26$\pm$0.01 & 16.00$\pm$0.01 & 15.77$\pm$0.02 & 15.45$\pm$0.04 & 15.14$\pm$0.09 & sdL0.5$^4$ & M7p & 65.09\\
12:56:37.10 & $\--$02:24:52.5 & 19.14$\pm$0.03 & 17.27$\pm$0.02 & 16.77$\pm$0.01 & 16.08$\pm$0.01 & 16.05$\pm$0.01 & $\--$ & 15.21$\pm$0.04 & 15.01$\pm$0.08 & sdL3.5$^5$ & M6p & 93.10\\
13:33:48.27 & +27:35:05.5 & 20.14$\pm$0.05 & 18.21$\pm$0.04 & 17.47$\pm$0.02 & 16.62$\pm$0.01 & 16.27$\pm$0.02 & 15.98$\pm$0.02 & $\--$ & $\--$ & sdL3$^6$ & M8p & 35.56\\
13:50:58.86 & +08:15:06.8 & 20.88$\pm$0.08 & 18.99$\pm$0.06 & 18.66$\pm$0.05 & 17.93$\pm$0.04 & 18.07$\pm$0.10 & 17.95$\pm$0.15 & 17.45$\pm$0.20 & $\--$ & sdL5$^7$ & M5.5p & 77.49\\
14:14:05.74 & $\--$01:42:02.7 & 19.55$\pm$0.03 & 17.97$\pm$0.03 & 17.50$\pm$0.03 & 16.81$\pm$0.02 & 16.45$\pm$0.03 & 16.14$\pm$0.03 & $\--$ & $\--$ & sdL0.5$^1$ & M7p & 23.45\\
14:16:24.08 & +13:48:26.7 & 18.00$\pm$0.02 & 15.36$\pm$0.02 & 14.26$\pm$0.01 & 12.99$\pm$0.01 & 12.47$\pm$0.01 & 12.05$\pm$0.01 & $\--$ & $\--$ & sdL7$^8$ & L4p & 152.37\\\hline\end{tabular}
\tablefoot{References to the spectroscopic classification:
$^1$\citet{espinoza13}, $^2$\citet{lodieu12},
$^3$\citet{kirkpatrick10}, $^4$\citet{lodieu12a},
$^5$\citet{sivarani09,subdwarf4}, $^6$\citet{zhang12}, $^7$\citet{lodieu10}, $^8$\citet{schmidt10}}
\label{subdwarfs_030915}
\end{table*}

\subsection{Peculiar blue sources, $Y-J<0.8$}
\label{peculiarblue}

Peculiar blue L and T dwarfs could be missed because they are
classified earlier than L0, or have $Y-J$ colours bluer than the
selection limit $Y-J=0.8$ (these are nearly the same thing). To
  investigate this issue we began by analysing the colours of known L
  and T subdwarfs. Subdwarfs however are only the most extreme
  examples of peculiar blue sources, being outnumbered by low
  metallicity members of the thick disk. To estimate the
  incompleteness due to blue L dwarfs being classified M, we use the
  sample of L dwarfs from \citet{schmidt10}, which is free of colour
  biases.

\subsubsection{Colours of subdwarfs}

We searched the literature for all subdwarfs for which we were
  able to collect $izYJHKW1W2$ photometry in at least 6 of the
  bands. The 9 subdwarfs satisfying these criteria are listed in Table
  \ref{subdwarfs_030915}. The table lists coordinates, photometry,
  spectral type (SpT), the {\em photo-type} classification (PhT), the
  $\chi^2$ of the fit, and the reference to the discovery paper. Seven
  of the nine sources have {\em photo-type} classifications earlier
  than L0, and of these, six have $Y-J<0.8$. None were considered
  before: five are recent discoveries and are not in
  DwarfArchives.org; one has $J>17.5$; one has not been observed in
  $K$ in UKIDSS. Of the other two sources, both with $Y-J>0.8$, the
  sdL7 \object{2MASS J11582077+0435014} (from \citealt{kirkpatrick10})
  is in our sample, and the sdL7 \object{SDSS J141624.09+134826.7}
  (from \citealt{schmidt10}) would be, but it is just brighter than
  our catalogue bright limit $J=13$. The bluer colours of L subdwarfs
  result in {\em photo-type} classifications that are on average
  between four and five spectral sub-types earlier than the spectral
  classification. This indicates that our sample will miss most
  subdwarfs of type sdL4 and earlier, but will include most subdwarfs
  of type sdL5 and later.

Table \ref{subdwarfs_030915} contains 6 objects in the range $13.0\le J\le17.5$,
and detected in $YJHK$, i.e. within our search volume, and therefore
amounting to $0.44\%$ of the LT population. This would be an
underestimate of the total proportion of subdwarfs in the LT
population, as the sample is incomplete. This limit $>0.44\%$ is
substantially larger than the figure favoured by \citet{chabrier03} of
$0.2\%$. The subdwarf fraction in the cool dwarf regime is
bounded to be $<0.68\%$, for all M stars \citep{covey08}, and
$>0.02\%$, for types $\ge$M5 \citep{lodieu12}, values consistent with
the estimate of \citet{chabrier03}. It is possible that not all the sources in Table
\ref{subdwarfs_030915} are genuine subdwarfs, but that some are
thick disk sources.

Table \ref{subdwarfs_030915} contains 5 sources $13.0\le J\le17.5$,
detected in $YJHK$, classified by {\em photo-type} as M, implying a
lower limit to the incompleteness to peculiar blue sources of $>0.37\%$.

Although the colours of subdwarfs lead to earlier classifications, the
$\chi^2$ values for the fits, listed in Table \ref{subdwarfs_030915},
are rather large, and all but one of the known subdwarfs would be
identified as peculiar, with $\chi^2>20$.  We therefore investigated
relaxing the $Y-J$ colour cut, attempting to identify subdwarfs as
objects with M-star {\em photo-type} classifications but with large
$\chi^2$. This proved unsuccessful, because the number counts of M
stars increase so steeply towards bluer colours, that the L subdwarfs
are greatly outnumbered by M stars with peculiar colours, and so
cannot be picked out. We also attempted to use the known subdwarfs to
define colour templates. This also failed, because the differences in
colour between subdwarfs of the same spectral type are as great as the
difference between a particular subdwarf and the nearest MLT template.

Although we have been unable to develop a method to identify complete and
uncontaminated samples of subdwarfs of type sdL4 and earlier, there
would be value in pursuing this problem further, as even a sample
with, say, $50\%$ contamination (requiring spectrosopic confirmation),
would provide a valuable complement to samples derived using proper
motion.

\subsubsection{Incompleteness estimate}

In this section we use the sample of 484 L dwarfs of
  \cite{schmidt10} to estimate the incompleteness due to peculiar blue
  L dwarfs being classified M. The sample of \cite{schmidt10} is
  particularly useful because of the lack of colour bias. A very blue
  cut in the colour $i-z$ was taken, sufficient to ensure inclusion of
  all L dwarfs.

We matched the L dwarf sample of \cite{schmidt10} to UKIDSS and WISE
and limited attention to the 142 sources with reliable photometry in
all bands $izYJHKW1W2$, and brighter than $J=17.5$. We classified this
sample using {\em photo-type} and then examined the colours and
classifications. Clipping outliers where the {\em photo-type} and
spectroscopic classifications differed by $\ge 3$ sub-types, we
measured a r.m.s. scatter in the {\em photo-type}
classifications of precisely 1.0 sub-types. This uncertainty agrees
with our previous estimate (Paper I). Therefore, except for outliers
differing by $\ge 3$ sub-types, our catalogue of L dwarfs is unbiased,
since we will account for this scatter in the calculation of the
luminosity function (see \S2). There are 3 sources for which the
classification differs by $\ge 3$ sub-types, and are classified M, and
there are a further two sources with colours $Y-J<0.8$. These 5
sources are therefore peculiar blue sources that are missed by {\em
  photo-type}. Of these, 3 are L0, 1 is L1 and 1 L3. Counting the
number of sources in these classes in the \cite{schmidt10} sample, and
applying this fractional incompleteness to the same bins in our sample
of 1361 sources, results in a computed incompleteness of $3\%$ due to
peculiar blue sources classified as M.
 
A corollary of the conclusion that {\em photo-type} classifications
for peculiar blue sources are biased towards earlier spectral types is
that the {\em photo-type} classifications for peculiar red sources
will be biased towards later spectral types. For example the
L2$\gamma$ dwarf \object{2MASS J01262109+1428057} \citep{metchev08},
previously noted, which has $\chi^2=202$, has a {\em photo-type}
classification of L7. We can expect our catalogue to be
correspondingly overcomplete for peculiar red objects, by including
sources classified as early L that are actually peculiar red M
stars. Such peculiar red sources are typically young, and therefore of
low mass, and may include examples with planetary masses. The
proportion of such sources is not known, but they should be found
among the sources with large $\chi^2$. Clearly a useful exercise would
be to obtain spectra of all 97 sources with $\chi^2>20$, to
characterise the peculiar blue and red populations and quantify their
numbers.

\section{Spectroscopic follow up of peculiar sources}
\label{spectra}

\begin{figure*} 
\includegraphics[width=18.0cm]{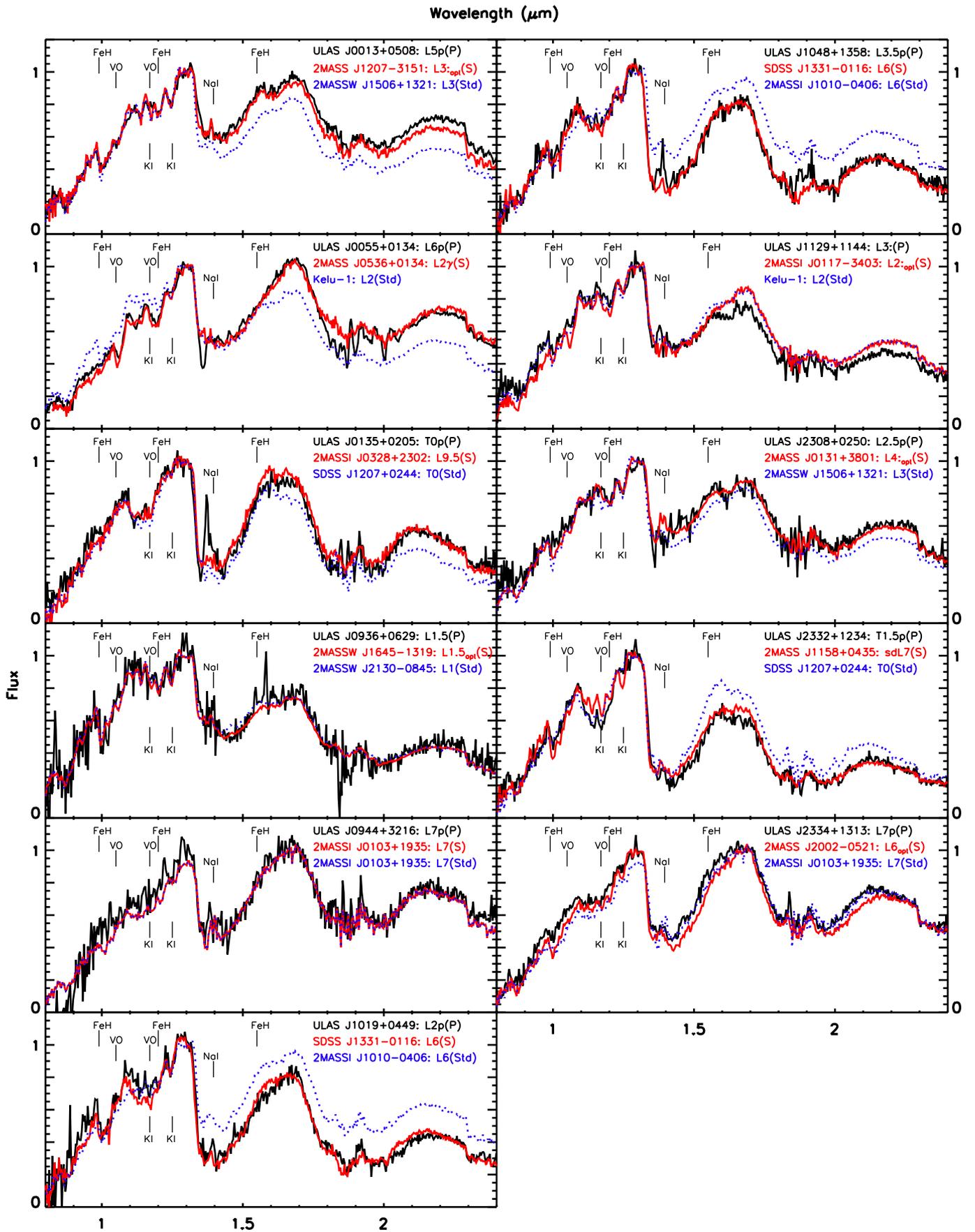} 
\caption{SpeX spectra (black) of the 11 sources observed. Overplotted
  are the best-fit spectroscopic standard spectrum (blue), and the best match
  spectrum from the SPL (red). Coordinates of the standards are provided in
  \citet{kirkpatrick10}. Coordinates of the SPL sources are provided in the text.}
\label{spexspectra} 
\end{figure*}

\begin{figure*} 
\includegraphics[width=16.0cm]{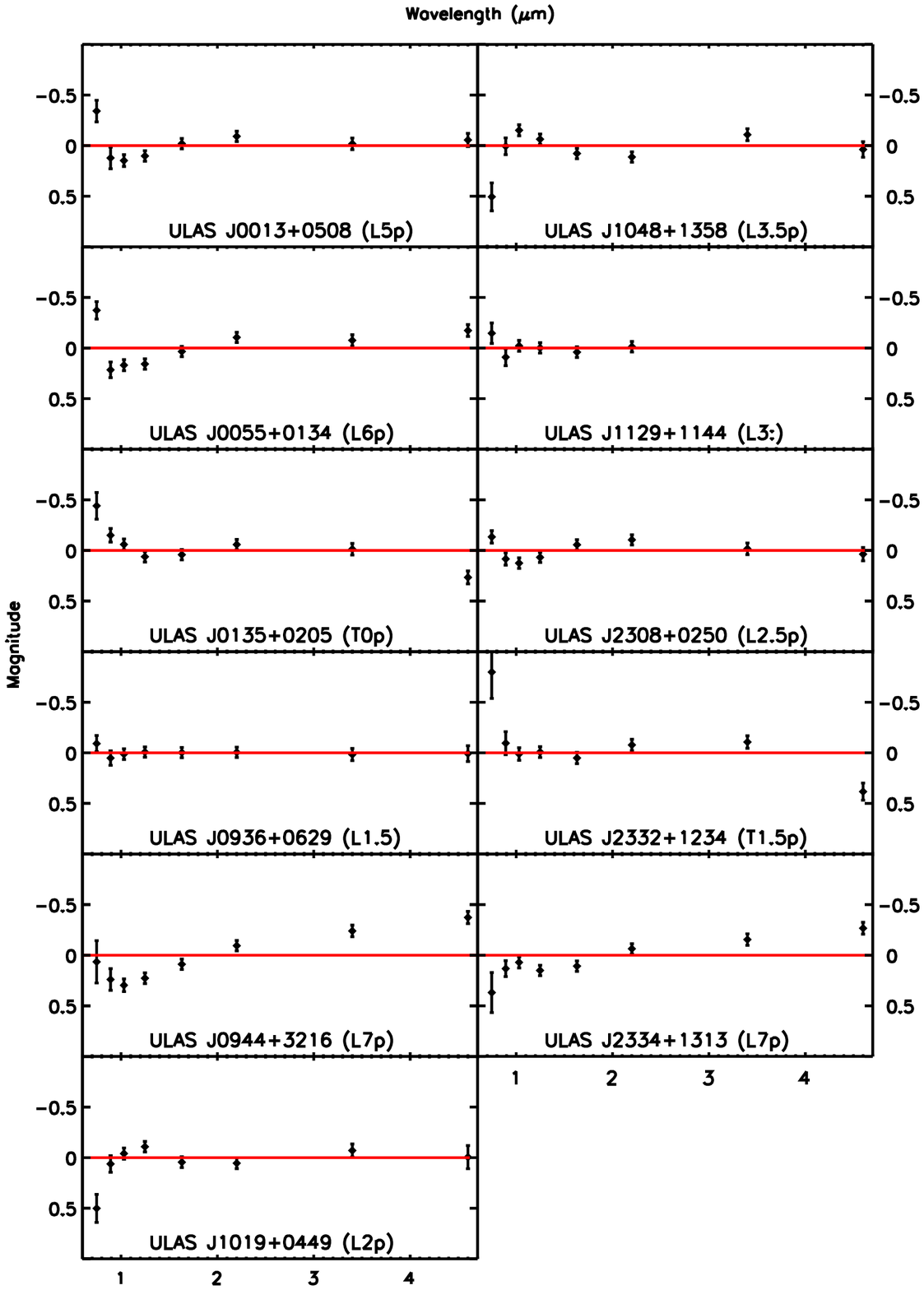} 
\caption{SEDs of the 11 sources observed. For each source the
  difference in mag. between the source SED and the (labeled) best-fit {\em photo-type} template is plotted. Points above the line correspond to the source being brighter than the template.}
\label{seds} 
\end{figure*}

In Paper I we presented spectra of 8 sources from the {\em photo-type}
LT catalogue. All had $\chi^2<20$, i.e. were classified as normal
rather than peculiar. All sources were confirmed as normal ultracool
dwarfs, and there was very close agreement between the spectroscopic
classification and the {\em photo-type} classification.

In the current paper we present spectra of 11 additional sources, of
which 9 have $\chi^2>20$, i.e. are classified as peculiar.  The
coordinates of the 11 sources are provided in Table
\ref{observing_details}, together with details of the observations. In
the following we use the short names provided in Table
\ref{observing_details}. The spectra were obtained with the SpeX
spectrograph mounted on the 3m NASA Infrared Telescope Facility (IRTF)
over several nights in March, May and November 2013.  Photometry of
the sources and the {\em photo-type} and spectroscopic classifications
are provided in Table \ref{fu_peculiar}. The spectra are presented in
Fig. \ref{spexspectra}. All sources are confirmed as ultracool dwarfs.

The conditions over the runs were variable with patchy clouds. The
seeing FWHM was in the range $0.8 -1.0$ $\arcsec$ at $K$. We operated in
prism mode with the 0.8$\arcsec$ slit aligned at the parallactic angle
and obtained low-resolution ($\lambda$/$\Delta$ $\lambda$ $\sim$90)
near-infrared spectral data spanning 0.7 - 2.5 $\mu$m. Each target
was first acquired in the guider camera. Exposure times varied from
150s to 180s depending on the brightness of the target. Six to 12
images were obtained for each object in an ABBA dither pattern along
the slit. An A0V star was observed immediately after each target at
similar airmass, for flux calibration and telluric
correction. Internal flat-field and Ar arc lamp exposures were
acquired for pixel response and wavelength calibration,
respectively. All data were reduced using SpeXtool version 3.3
(\citealt{vacca03}, \citealt{cushing04}) using standard settings, with
the exception that we modified the procedure to correct for telluric
absorption, by accounting for the difference between the airmass the
target was observed at and the airmass the telluric standard was
observed at.

Spectral types were first estimated by visually comparing each object
to the near infrared spectral standards from
\citet{kirkpatrick10}. All spectra were normalized as described in
\citet{kirkpatrick10} and the best fit was determined by
eye. Subsequently, we compared each spectrum to the library of
optically classified M-T dwarfs in the SpeX Prism Library (SPL;
\citet{burgasser14}) and applied a chi-square minimisation routine to
determine the closest object match (see \citet{cushing08} for
technique description). Fig. \ref{spexspectra} shows the spectrum with its
{\em photo-type} (black), the best visual
spectral standard match (blue), and the best fit object from the SPL with
its optical spectral type displayed (red).

Of the 11 sources with spectra, 2 objects have $\chi^2$ $<$ 20, i.e.
were phototyped as normal.  Our spectral analysis confirms that ULAS
J0936+0629 with $\chi^2$=2 is a field L1 and ULAS J1129+1144 with
$\chi^2$=4 is a field L2. These classifications are very similar to
the {\em photo-type} classifications of L1.5 and L3:, respectively, as
expected.

Nine objects in the spectral sample have $\chi^2$ $>$ 20, i.e. were
phototyped as peculiar.  We plot the difference in mag. in each band
between the object SED and the best-fit {\em photo-type} template in
Fig. \ref{seds} so that it is possible to see which wavelengths
contribute most to the large $\chi^2$. In this plot, points above the
line correspond to the source being brighter than the template.  Our
spectral analysis yields the following:

{\bf ULAS J0013+0508} The spectrum (Fig. \ref{spexspectra}) is best
fit by the optical L3 \object{2MASS J12070374-3151298}
(\citealt{burgasser10}) and the L3 IR standard from
\citet{kirkpatrick10}. The source shows no obvious peculiar
  spectral features in the near-infrared region, except that it is
  redder than the standard. The large $\chi^2$=25.8 arises because the
  source is red in $Y-K$ but blue in $i-z$ compared to the template
  colours. An optical spectrum would be useful to confirm this
  mismatch, as this may highlight the strength of {\em photo-type}
  compared to spectral classification i.e. the ability to identify
  peculiar objects because of the large wavelength coverage. The
  source ULAS J2308+0250 displays similar behaviour.

{\bf ULAS J0055+0134} We find the spectrum of ULAS J0055+0134 shows
classic signatures of low-surface gravity with deep VO absorption and
weak alkali line absorption (e.g. \citealt{allers13}).  The best
(poorly) fit IR standard from \citet{kirkpatrick10} is Kelu-1 (L2)
although ULAS J0055+0134 is significantly redder, a trademark of young
low-gravity brown dwarfs (e.g. \citealt{faherty12, faherty13}).  Using
the \citet{allers13} indices which take into account VO, FeH,
continuum, and KI, we find this source would be considered a very
low-gravity (VL-G) L2.  The best fit individual object is the optical
L2$\gamma$ 2M0536+0134 (Faherty et al., in prep, where coordinates
will be provided), which is in agreement with the IR indices
evaluation, as the $\gamma$ sources on the \citet{Cruz09} optical
classification scheme are also very low gravity and likely younger
than the Pleiades ($<$ $\sim$ 120 Myr, \citealt{stauffer98}).

The $\chi^2$=60.8 correctly identifies this source as peculiar and the
{\em photo-type} prediction of L6p is incorrect but logical given how
red the young brown dwarfs are compared to field sources.

{\bf ULAS J0135+0205} The spectrum (Fig. \ref{spexspectra}) is best
fit by the L9.5 \object{2MASSI J0328426+230205} (\citealt{burgasser08}) and the T0 IR
standard from \citet{burgasser06b}. The {\em photo-type} class is T0p, and the
large $\chi^2$=37.9 is evident in Figure \ref{seds}, where the SED is
highly discrepant compared to the T0 template, being too blue from $i$
to $J$, and again in $W1-W2$. Since the near infrared spectrum (and
photometry) appears normal, an optical or mid infrared spectrum would
aid in deciphering the potential peculiarity.

{\bf ULAS J0944+3216} The spectrum (Fig. \ref{spexspectra}) is best
fit by the L6$\beta$ \object{2MASSI J0103320+193536}
(\citealt{cruz04}, \citealt{allers13}) which happens to be the L7 near
infrared standard from \citet{kirkpatrick10}.  \citet{allers13} give
\object{2MASSI J0103320+193536} an intermediate gravity
classification, hence this is a mildly low surface gravity brown
dwarf. ULAS J0944+3216 shows similar signs of an intermediate gravity
due to its H band continuum.  However low surface gravity spectral
deviations are difficult to pinpoint beyond L5. Therefore this
designation should be considered tentative.

The {\em photo-type} class is L7p with very large $\chi^2$=106.9,
evident in Figure \ref{seds} where there is a mismatch to the template
at nearly all wavelengths.  Some of this may be attributed to the
apparent general mismatch of the templates at spectral type L7, over
the interval $Y-K$, noted before, and visible in
Fig. \ref{colour3}. This further emphasises the importance of
investigating the discontinuity in $Y-K$ between spectral types L7 and
L8.

{\bf ULAS J1019+0449 and \bf ULAS J1048+1358} The spectra of both ULAS
J1019+0449 and ULAS J1048+1358 are best (poorly) fit by the near
infrared L6 standard from \citet{kirkpatrick10}.  The best match from
the SPL is the unusually blue high $v_{tan}$ source \object{SDSS
  J133148.92-011651.4} (\citealt{faherty09}, \citealt{kirkpatrick10},
\citealt{burgasser10}).  Similarly both sources received a {\em
  photo-type}class of L2p/L3.5p, much earlier than the spectral best
fits.  As expected these objects are unusually blue for their types
which drove the {\em photo-type} class to an earlier but peculiar
classification.  Assuming a spectrophotometric distance using the
\citet{dupuy12} relations along with the proper motion extracted from
multiple epochs, $0.5\arcsec$/yr, we find that ULAS J1019+0449 has a
moderately high $v_{tan}$ ($\sim$ 85 km s$^{-1}$).  The measured
proper motion of ULAS J1048+1358 is $<0.1\arcsec$/yr.  We conclude that
both sources warrant the peculiar designation allotted by {\em
  photo-type} and implied by their large $\chi^{2}$ values.

Interestingly ULAS J1019+0449 has an enticingly sharply peaked $H$ band
that typically identifies young low surface gravity brown dwarfs.
However \citet{aganze16} have recently shown that low
metallicity, high surface gravity brown dwarfs can mimic this same
feature with a reason likely linked to condensation efficiency or
changes in the collision induced H$_2$ absorption.

{\bf ULAS J2308+0250} The spectrum (Fig. \ref{spexspectra}) is best
fit by the optical L4 \object{2MASS J01311838+3801554}
(\citealt{burgasser10}) and the L3 IR standard from
\citet{kirkpatrick10}.  The {\em photo-type} of L2.5p with mild
$\chi^2$=20 is consistent with the best fits.  The source shows no
obvious peculiar spectral features. The SED mismatch of this source is quite
similar to that of ULAS J0013+0508 i.e. red in the near-infrared and
blue in the optical. An optical spectrum would be useful to confirm
this mismatch.

{\bf ULAS J2332+1234} The spectrum (Fig. \ref{spexspectra}) is best
(poorly) fit by the T0 IR standard from \citet{burgasser06b}.  The
sdL7 \object{2MASS J11582077+0435014} from \citet{kirkpatrick10} is
the best fit to the spectrum of ULAS J2332+1234.  Compared to the T0
IR standard displayed \---\ which was the best standard match
\---\ ULAS J2332+1234 shows supressed $H$ and $K$ bands, a hallmark of
low-metallicity objects.  Compared to the known sdL7, ULAS J2332+1234
does not demonstrate the same depth of FeH absorption though it
matches well in $H$ and $K$.  The {\em photo-type} of T1.5p with
$\chi^2$=36.4 is consistent with low-metallicity peculiar sources
being typed later given their blue SEDs. We measure a proper motion of
$0.4\arcsec$/yr for this source, which adds to the evidence of low
metallicity.

{\bf ULAS J2334+1313} Similarly to ULAS J0944+3216, the {\em
  photo-type} class L7p and the spectral class L7 agree for this
source.  In this case, the best fit from the SPL was the L6 field
object \object{2MASS J20025073-0521524}
(\citealt{burgasser08}). Unlike ULAS J0944+3216, the source does not
fall on the intermediate gravity scale based on its $H$ band
continuum. The SED mismatch, with $\chi^2=50.9$, is less severe than
for ULAS J0944+3216. This source may be further evidence for the
problematic typing of L7 objects discussed in section 3.

\begin{table}\scriptsize
\centering
\caption{Observing details of the spectroscopic observations}
\begin{tabular}{c c c r r}
\hline\hline
Name & Short & {Date (UT)} & {$t_{exp}$} &  {A0 star}\\
     & name  &  & sec. &  HD no. \\\hline
ULAS J001306.33+050851.2 & ULAS J0013+0508 & 24/11/2013 & 1800 & 219833\\
ULAS J005505.69+013436.0 & ULAS J0055+0134 & 24/11/2013 & 1200 & 13936\\
ULAS J013525.37+020518.5 & ULAS J0135+0205 & 20/11/2013 & 1500 & 13936\\
ULAS J093621.87+062939.1 & ULAS J0936+0629 & 20/11/2013 & 1500 & 71908\\
ULAS J094419.56+321605.2 & ULAS J0944+3216 & 27/03/2013 & 1800 & 89239\\
ULAS J101950.97+044941.0 & ULAS J1019+0449 & 24/11/2013 & 1080 & 89239\\
ULAS J104814.77+135832.7 & ULAS J1048+1358 & 22/11/2013 & 1440 & 89239\\
ULAS J112926.00+114436.2 & ULAS J1129+1144 & 10/05/2013 &  800 & 97585\\
ULAS J230852.99+025052.0 & ULAS J2308+0250 & 20/11/2013 & 1500 & 219833\\
ULAS J233227.03+123452.1 & ULAS J2332+1234 & 24/11/2013 & 1800 & 210501\\
ULAS J233432.53+131315.3 & ULAS J2334+1313 & 24/11/2013 & 1500 & 210501\\\hline
\end{tabular}
\label{observing_details}
\end{table}

% % % % % % % % % % % % % % % % % % % % % % % % % % % % % % % % % % % % % % 
\begin{table*}\scriptsize
\caption{Photometry and spectral types of the 11 sources observed spectroscopically}
\begin{tabular}{c c c c c c c c c l r l}
\hline\hline
Name & {$i\pm \sigma_i$}&{$z\pm
  \sigma_z$}&{$Y\pm \sigma_Y$}&{$J\pm \sigma_J$}&{$H\pm
  \sigma_H$}&{$K\pm \sigma_K$}&{$W1\pm \sigma_{W1}$}&{$W2\pm \sigma_{W2}$}&{PhT}&{$\chi^2$}&{SpT}\\\hline
ULAS J0013+0508 & 20.91$\pm$0.09 & 19.04$\pm$0.09 & 18.10$\pm$0.03 & 16.75$\pm$0.02 & 15.72$\pm$0.01 & 14.91$\pm$0.01 & 14.33$\pm$0.03 & 13.97$\pm$0.04 & L5p & 25.8 & L3\\
ULAS J0055+0134 & 20.67$\pm$0.07 & 18.75$\pm$0.06 & 17.71$\pm$0.02 & 16.37$\pm$0.01 & 15.29$\pm$0.01 & 14.40$\pm$0.01 & 13.71$\pm$0.03 & 13.25$\pm$0.03 & L6p & 60.8 & L2$\gamma$\\
ULAS J0135+0205 & 21.76$\pm$0.12 & 18.69$\pm$0.05 & 17.55$\pm$0.02 & 16.48$\pm$0.02 & 15.66$\pm$0.01 & 14.99$\pm$0.01 & 14.28$\pm$0.03 & 13.88$\pm$0.04 & T0p & 37.9 & T0\\
ULAS J0936+0629 & 20.44$\pm$0.06 & 18.55$\pm$0.05 & 17.62$\pm$0.02 & 16.45$\pm$0.01 & 15.76$\pm$0.01 & 15.15$\pm$0.01 & 14.81$\pm$0.03 & 14.53$\pm$0.06 & L1.5 & 2.0 & L1\\
ULAS J0944+3216 & 22.12$\pm$0.20 & 19.59$\pm$0.09 & 18.61$\pm$0.04 & 17.19$\pm$0.02 & 16.08$\pm$0.01 & 15.15$\pm$0.01 & 14.24$\pm$0.03 & 13.69$\pm$0.03 & L7p & 106.9 & L7\\
ULAS J1019+0449 & 21.55$\pm$0.13 & 19.06$\pm$0.07 & 18.06$\pm$0.02 & 16.82$\pm$0.01 & 16.24$\pm$0.02 & 15.63$\pm$0.02 & 15.11$\pm$0.04 & 14.89$\pm$0.10 & L2p & 21.4 & L6p\\
ULAS J1048+1358 & 21.61$\pm$0.13 & 18.97$\pm$0.07 & 17.88$\pm$0.02 & 16.72$\pm$0.01 & 16.04$\pm$0.01 & 15.38$\pm$0.01 & 14.64$\pm$0.03 & 14.50$\pm$0.06 & L3.5p & 33.4 & L6p\\
ULAS J1129+1144 & 20.66$\pm$0.09 & 18.80$\pm$0.07 & 17.76$\pm$0.02 & 16.56$\pm$0.01 & 15.81$\pm$0.02 & 15.08$\pm$0.02 &  $\--$         &  $\--$         & L3: & 4.0 & L2\\
ULAS J2308+0250 & 20.06$\pm$0.04 & 18.21$\pm$0.04 & 17.34$\pm$0.02 & 16.08$\pm$0.01 & 15.20$\pm$0.01 & 14.50$\pm$0.01 & 14.16$\pm$0.03 & 13.92$\pm$0.04 & L2.5p & 20.0 & L3\\
ULAS J2332+1234 & 22.30$\pm$0.26 & 19.37$\pm$0.10 & 18.10$\pm$0.04 & 16.90$\pm$0.02 & 16.39$\pm$0.03 & 15.88$\pm$0.03 & 15.18$\pm$0.04 & 14.77$\pm$0.07 & T1.5p & 36.4 & T0\\
ULAS J2334+1313 & 21.91$\pm$0.19 & 18.96$\pm$0.06 & 17.86$\pm$0.02 & 16.60$\pm$0.01 & 15.58$\pm$0.01 & 14.66$\pm$0.01 & 13.80$\pm$0.03 & 13.28$\pm$0.03 & L7p & 50.9 & L7\\\hline
\end{tabular}
\label{fu_peculiar}
\end{table*}

\subsection{Candidate binaries}

In Paper I, \S2.4, we described a method to identify candidate
unresolved LT binaries using the SEDs. The method identifies sources
where the SED fit is improved significantly for a binary, compared to
a single source. The method is more or less effective depending on the
two dwarfs comprising the binary, and was shown to be particularly
sensitive for the case of a primary near L5 and a secondary near
T5. Three of the sources listed in Table \ref{observing_details} were
identified as candidate binaries: ULAS J0135+0205 (L6+T3), ULAS
J1019+0449 (L3+T4), and ULAS J2332+1234 (L8+T4). For the second and
third of these, the spectra indicate that the large $\chi^2$ of the
{\em photo-type} fit is a consequence of the low metallicity of the
source, rather than due to binarity. For ULAS J0135+0205 we find that
over the near-infrared region the spectrum is better fit as single
(T0), rather than by the binary combination ((L8+T4) suggested by the
photometry. On the evidence of these three sources it appears that the
method more often picks out single sources with spectral peculiarities
that mimic the colours of binaries, rather than picks out genuine
unresolved binaries.

\section{Summary}
\label{summary}

In this paper we presented a large sample of ultracool dwarfs, with
accurate spectral types, comprising 1281 L dwarfs and 80 T dwarfs,
brighter than $J=17.5$. This is the first large homogeneous sample
covering the entire spectral range L0 to T8, and will be valuable for
statistical studies of the properties of the population, including
measuring the substellar mass function, measuring the disk scale
height (in conjunction with a deeper sample), and quantifying the
spread in metallicity and surface gravity in the population. Because
the sample is large it will also be useful for identifying rare types
of L and T dwarf, including young red sources, and identifying
benchmark systems.

\begin{acknowledgements}
We are grateful to Sarah Schmidt for pointing out the issue with the
$i-z$ colours of our M star templates, and for help in improving the
templates. We thank Paul Hewett for providing new quasar colour
templates, extended to include W1 and W2, and Mike Read for help in understanding the
details of the UKIDSS matching algorithm. We are also grateful to
Daniella Bardalez Gagliuffi for useful discussions related to the
spectral typing of objects.\\
We are grateful to the anonymous referee for a very detailed report
that resulted in substantial improvements to the paper.\\
This research has benefited from the SpeX Prism Spectral Libraries,
maintained by Adam Burgasser at http://pono.ucsd.edu/~adam/browndwarfs/spexprism.\\
This publication makes use of data products from the Wide-field
Infrared Survey Explorer, which is a joint project of the University
of California, Los Angeles, and the Jet Propulsion
Laboratory/California Institute of Technology, and NEOWISE, which is a
project of the Jet Propulsion Laboratory/California Institute of
Technology. WISE and NEOWISE are funded by the National Aeronautics
and Space Administration.\\
This research has benefited from the M, L,
T, and Y dwarf compendium housed at DwarfArchives.org. The UKIDSS
project is defined in \citet{UKIDSS}. UKIDSS uses the UKIRT Wide Field
Camera \citep{casali07}. The photometric system is described in
\citet{hewett06}, and the calibration is described in
\citet{hodgkin09}. The science archive is described in
\citet{hambly08}.\\
Funding for SDSS-III has been provided by the Alfred P. Sloan
Foundation, the Participating Institutions, the National Science
Foundation, and the U.S. Department of Energy Office of Science. The
SDSS-III web site is http://www.sdss3.org/.  SDSS-III is managed by
the Astrophysical Research Consortium for the Participating
Institutions of the SDSS-III Collaboration including the University of
Arizona, the Brazilian Participation Group, Brookhaven National
Laboratory, Carnegie Mellon University, University of Florida, the
French Participation Group, the German Participation Group, Harvard
University, the Instituto de Astrofisica de Canarias, the Michigan
State/Notre Dame/JINA Participation Group, Johns Hopkins University,
Lawrence Berkeley National Laboratory, Max Planck Institute for
Astrophysics, Max Planck Institute for Extraterrestrial Physics, New
Mexico State University, New York University, Ohio State University,
Pennsylvania State University, University of Portsmouth, Princeton
University, the Spanish Participation Group, University of Tokyo,
University of Utah, Vanderbilt University, University of Virginia,
University of Washington, and Yale University.
\end{acknowledgements}
\bibliographystyle{aa}
\bibliography{Skrzypek-PaperII}

\end{document}